\documentclass[sigconf,balance=false]{acmart}
\usepackage{popets}

\acmISBN{}
\acmConference{}

\newtheorem{lemma}{Lemma}

\usepackage[ruled,vlined,linesnumbered]{algorithm2e}
\usepackage{algorithmicx}
\usepackage[noend]{algpseudocode}
\usepackage{algcompatible}

\usepackage{amsmath,amsfonts}

\usepackage{amssymb}

\usepackage{bm, mathtools, extpfeil}

\usepackage{xcolor}
\usepackage{comment}

\usepackage{makecell}
\usepackage{hhline}
\usepackage{multirow}
\usepackage{cellspace}
\usepackage{float}
\usepackage{url}
\usepackage{wrapfig}
\usepackage{graphicx}
\usepackage{pgfplots}
\usepackage{subfloat}
\usepackage{epstopdf}
\usepackage{marginnote}
\usepackage{nameref}
\usepackage{enumitem}
\usepackage{multirow}
\usepackage{booktabs}
\usepackage{makecell}
\usepackage{caption}

\newcommand{\fR}{\mathsf{R}}

\newcommand{\LWE}{\mathsf{LWE}}
\newcommand{\RLWE}{\mathsf{RLWE}}

\newcommand{\ring}[3]{\mathcal{R}^{#2}_{#3}{#1}}
\newcommand{\prlwe}[3]{\mathcal{P}^{#1}_{#3}{(#2)}}

\newcommand{\round}[1]{\left\lfloor #1 \right\rceil} 

\newcommand{\dom}{\mathsf{Dom}}
\newcommand{\img}{\textsf{Img}}

\renewcommand{\Im}{\mathsf{Im}}
\renewcommand{\Re}{\mathsf{Re}}

\renewcommand{\vec}[1]{\mathbf{#1}}

\usepackage{color}

\begin{document}

\title[TETRIS: Composing FHE Techniques for Private Functional Exploration Over Large Datasets]{TETRIS: Composing FHE Techniques for Private Functional Exploration Over Large Datasets}


\author{Malika Izabachène}
\orcid{0000-0003-0216-7958}
\affiliation{
    \institution{Independent}
    \city{Paris}
    \country{France}
}
\email{malika.izabachene@gmail.com}

\author{Jean-Philippe Bossuat}
\orcid{0000-0002-2020-0224}
\affiliation{
    \institution{Independent}
    \city{Lausanne}
    \country{Switzerland}
}
\email{jeanphilippe.bossuat@gmail.com}

\renewcommand{\shortauthors}{}

\begin{abstract}
To derive valuable insights from statistics, machine learning applications frequently analyze substantial amounts of data. In this work, we address the problem of designing efficient secure techniques to probe large datasets which allow a scientist to conduct large-scale medical studies over specific attributes of patients' records, while maintaining the privacy of his model. We introduce a set of composable homomorphic operations and show how to combine private functions evaluation with private thresholds via approximate fully homomorphic encryption. This allows us to design a new system named TETRIS, which solves the real-world use case of private functional exploration of large databases, where the statistical criteria remain private to the server owning the patients’ records. 
Our experiments show that TETRIS achieves practical performance over a large dataset of patients even for the evaluation of elaborate statements composed of linear and nonlinear functions. It is possible to extract private insights from a database of hundreds of thousands of patient records within only a few minutes on a single thread, with an amortized time per database entry smaller than $2$ms.
\end{abstract}


\maketitle

\section{Introduction}
\label{section:intro}
Large-scale models are used for several tasks such as DNA research, for understanding and generating textual contents, or producing code and many of them are trained on data extracted from millions of gigabytes text words. 
While these large models are significant across multiple applications, their deployment sets serious privacy challenges as both the user's requests and the model service provider are exposed to sensitive information leak. The user usually wants to protect the privacy of its request to the large model service provider as they may contain sensitive information about the data. And the large model owner usually would like to protect the model intellectual property of his algorithms. In this paper, we consider the scenario where a scientist would like to conduct a medical study over a large dataset of sensitive patients’ data owned by a server. 
In this setting, the statistics run by the scientist are built from an intelligent system model that do contain intellectual property the scientist would like to protect. 
Our solution makes it possible for the scientist to probe a medical server in order to learn about public health while disclosing as little information as possible to the server. In particular, the response to the scientist request is kept encrypted on the server side and remains unknown to him. Moreover, the scientist does not learn more information on the datasets that what can be inferred from the response to his request. We explicit the function model owned by the scientist as a combination of specific attributes which defines the type of properties the scientist is probing. Such property could search how many selected patients’ profiles do have a blood sugar higher than some threshold. In that case, the selection criteria could consider some additional risk factors such as functional neurological disorders or/and blood pressure for example. While the application we focus on raises a medical use case, our framework could still have other application such as credit scoring where a scientist would like to run a private system model to classify individuals’ profiles according to confidential attributes.

\vspace{-1.8em}

\paragraph{FHE computation challenges for large models} One solution to the privacy risk of large models deployments is using \emph{fully homomorphic encryption} (FHE), which enables the evaluation of arbitrary functions over encrypted data. 
Using FHE as a privacy technology in the context of large datasets poses significant performance challenges, as the homomorphic operations become more costly as the number of data grows. Some FHE schemes support SIMD arithmetic which allows to pack data and perform the evaluation on many slots in parallel. However, many privacy-preserving applications involve different types of operations, some of which behave more or less nicely depending on the message encoding format. 
For example, in the medical use case study we consider in this work, the scientist must perform both nonlinear functions, such as combination of threshold functions, and linear functions such as scoring functions. 

The most known FHE schemes encode the message as a scalar or as a polynomial and rely on the Learning with Errors (LWE) or ring LWE (RLWE) assumptions respectively to provide IND-CPA security, meaning that the ciphertext remains indistinguishable to a random element in the ciphertext space. These FHE schemes perform the computation differently depending on the ciphertexts format and the optimization techniques available. LWE-based schemes support very efficient bit-wise computations which allows to manipulate and combine encrypted ciphertexts with high flexibility while RLWE-based schemes generally support SIMD computation and allows to efficiently process a large amount of data at once but leaves less margin to manipulate each ciphertext independently. The authors of \cite{SP:LHHMQ21} proposed a method to evaluate both a polynomial and a non-polynomial function but at the cost of introducing approximation errors in the calculation, which is not fully suitable for use cases that do not tolerate computation failures. 

From the functional standpoint, the first category of schemes has the advantage of supporting arbitrary circuit evaluations in a more efficient way.
The FHEW-like bootstrapping based schemes~\cite{EC:DucMic15,EC:GINX16,AC:CGGI16}, to cite a few, falls in this category and has shown to be able to evaluate discretized arbitrary functions at a slight additional cost compared to the gate bootstrapping, which allows to homomorphically reduce the noise induced by homomorphic computations in a FHE ciphertext. This generalized bootstrapping is called \emph{functional bootstrapping} or \emph{programmable bootstrapping} in the literature.
One of the major limitations of the functional bootstrapping technique is that if one wants to keep it efficient, the precision of the message encoding functions that one would like to evaluate needs to be restricted to relatively small sizes. For example, functions with domains larger than 8-to-10 bits of precision become difficult to evaluate in a reasonable amount of time. 
The second category of schemes does not provide generic methods to evaluate non-polynomial functions and alternatively needs to find efficient tailor-made polynomials to well approximate the function to be evaluated as shown in~\cite{LLNKSign2020, EC:LLKKNK22, TIFS:CKP22, LHDNN23}. 

\paragraph{Switching between the encodings in practice} Conversions between FHE ciphertexts format has raised interest over the last years~\cite{AC:CGGI17,JMC:BGGJ20,ACNS:CDKS21,SP:LHHMQ21,C:BCKPS23} as it allows to switch between LWE ciphertexts and RLWE ciphertexts back and forth, which leaves the possibility to take advantage of both format properties. One of the main procedures used to perform these conversion is homomorphic \emph{packing} (resp. \emph{repacking}) which allows to pack several LWE (resp. RLWE) ciphertexts into one RLWE ciphertext. 
Packing techniques can provide very low amortized cost for regular homomorphic operations such as additions, multiplications, and bootstrapping operations, compared to repeated invocations on individual ciphertexts, especially when the number of data points to be processed is huge.
These are well-suited for schemes whose modulus-to-noise ratio is super-polynomial.
Packing has also been proposed in the context of bootstrapping for polynomial modulus-to-noise ratio schemes using different techniques but as these schemes provide extremely efficient bootstrapping on individual ciphertexts, the amortized cost outperforms the regular bootstrapping cost only asymptotically. 
The authors of~\cite{JMC:BGGJ20} proposed a framework that allows to support arbitrary functions evaluations and SIMD arithmetic by switching from one format ciphertext representation to another. 
Their work reports timings of 7$s$ to switch from $N$ $\LWE$ ciphertexts to one $\RLWE$ ciphertext for $N=2^{12}$ and requires large public switching keys. 
The work from~\cite{ACNS:CDKS21} reports close to 1$s$ to pack a relatively small amount of data but both methods using a different packing method, but~\cite{JMC:BGGJ20} and \cite{ACNS:CDKS21} become impractical as the number of input ciphertexts grows.

The packing method from \cite{SP:LHHMQ21} takes advantage of the SIMD encoding and reports improved throughput compared to the previous state-of-the-art packing procedures but their method has remained relatively inefficient in term of amortized cost because it requires to evaluate a homomorphic modular reduction.
A nice improvement was recently proposed in~\cite{C:BCKPS23} for approximate homomorphic encryption.
Their experiments report a latency of $10.2$s to pack $2^{15}$ $\LWE$ ciphertexts in one $\RLWE$ ciphertext compared to $52$s to pack $2^{12}$ $\LWE$ ciphertexts using the method from~\cite{SP:LHHMQ21}.
However their method makes use of the LWE assumption over modules (Module LWE), which is not readily available in FHE libraries.
Also, once the packing is performed, it remains to combine SIMD arithmetic techniques and arbitrary functions evaluation to enable further homomorphic evaluations over hybrid circuits.
At the current state, there is still no fully satisfactory practical solutions that handle different types of operations including arbitrary functions and SIMD arithmetic over large domain without affecting the precision. 
\paragraph{Techniques for private functions evaluation over large datasets} Other cryptographic techniques for privacy-preserving computations have been previously proposed.
Multi-party computation (MPC) protocols have been proposed to solve the privacy issue in machine learning use cases (\cite{NDSS:SPTFBS21,bioinf:SKSSTKHL22, SPw:GM0SWY23,SP:MWAPR23,USENIX:LiuXieYu24} to mention a few of them) such as neural network evaluation, privacy-preserving models training, privacy-preserving record linkage (for large database). 
Most of them focus either on protecting the privacy of the datasets or/and the privacy of the function. 
The authors of~\cite{NDSS:SPTFBS21} proposed a solution for neural network training that instantiates an MPC framework using FHE to protect the privacy of the training data, the model and the evaluation. Their experiment processes 60K samples that are distributed among 10 parties. 
MPC techniques can also be used for two-party computation use cases, but we are not aware of an MPC-based work that solves the challenges addressed by our use case i.e. protecting the model functions while a possibly complex homomorphic computation is processed on a huge amount of data. 
Another advantage of FHE over traditional MPC techniques is that when ordering and combining homomorphic operations properly, the communication between both parties requires less interactions. 
In addition, our HE-based solution can be extended to support a partitioned database with only a linear cost in the setup phase (key-generation), which is not the case for traditional MPC-based solutions. 
The above privacy preserving techniques aim to protect the privacy of the database and/or function during the computation but cannot and are not designed to protect against attacks such as model inversion or individual singling out. 
In such cases, techniques such as differential privacy can be added on top of the FHE or MPC techniques to mitigate these attacks.
However, their use has some limitations as they inevitably incur a trade-off between utility and privacy.

\paragraph{Our contributions} 
\begin{itemize}
\item In this work, we design a new efficient solution named TETRIS allowing a scientist to probe a dataset containing a huge amount of data owned by a server. Our framework makes it possible for the scientist to privately request attribute functions over the data while maintaining the privacy of his attribute functions. On the other hand, the scientist does not learn more information on the patients’ data than what can be deduced from the response from the server.

\item We propose an adaptation of the large domain homomorphic evaluations of arbitrary functions technique from~\cite{PoPETS:IIMP22,FHEorg:BosIza24} in the case where the function is encrypted and the large dataset remains in the plaintext domain. In other words, the database owner processes the computation for the scientist who owns the function, and the large database doesn't need to be sent at any point. At the end, the database owner does not learn any information about the function that it evaluates while the scientist learns nothing except what could be inferred from the result and the computation circuit if the latter is known to him. Adapting the techniques from~\cite{PoPETS:IIMP22,FHEorg:BosIza24} allows us to overcome the domain size limitation constrained by using the functional bootstrapping. Our method enables to support the evaluation of arbitrary functions over large domains in an efficient way while keeping the functions private. In our framework, the arbitrary functions are instantiated to enable the homomorphic evaluation of scoring functions parametrized by the attributes. Our methodology for the evaluation of arbitrary function is introduced in Section~\ref{sections:large-domain} and its specialization to homomorphic scoring function evaluation is presented in Section~\ref{section:circuit}.

\item In the homomorphic thresholds evaluation phase, we make use of several homomorphic building blocks, including ring repacking, ring merging and ring switching, that we carefully combine to be able to merge and switch to a larger ring dimension. By doing so, we manage scheme switching via bootstrapping, and the evaluation of two consecutive nonlinear functions. In our framework, these nonlinear functions are specialized as a local and global threshold applied on linear combinations of encrypted inputs. Based on the approach from~\cite{LLNKSign2020} to evaluate the sign and the step functions over encrypted real inputs, we show how to evaluate threshold functions over encrypted discrete values with approximate homomorphic encryption. 
All these homomorphic operations are processed by thoroughly choosing appropriate dimensions and moduli to not affect precision, efficiency and security.

\item We implement a real-world use case that showcases how the above techniques can be combined together, along with approximate homomorphic encryption, to provide an efficient private database exploration framework, answering the following question: are there at least $x$ elements in the database that match at least $y$-out-of-$z$ private criteria? We show that such a question can be answered within a few minutes on a single thread for a database of hundreds of thousands of entries with an amortized cost of only $2ms$.

\item We tune the parameters involved in the noise growth of the underlying FHE operations, the precision, the correctness and the security of the homomorphic operations to propose an efficient implementation of each building block. We provide an open-source implementation for all aspects of this work using the Lattigo library~\cite{lattigo}.
\end{itemize}

\paragraph{Application to private database exploration} Assume a scientist would like to privately explore a database of size $p\times h$ to get a binary answer on whether this database contains a minimal set of items meeting private criteria of his choosing.
At a high level, this can be done by (i) evaluating a private scoring function on each of the $h$ attributes of each $p$ entry of the database, (ii) summing the encrypted scores to produce a single score for each $p$ entries of the database, (iii) evaluating a private threshold on the summed scores that returns $1$ if the score is larger than a private threshold, else $0$ and (iv) summing all binary values from the previous step to produce an encrypted count of how many entries match the private criteria and (v) evaluating a second private global threshold that returns $1$ if enough entries meet the private criteria, else $0$.

The private criteria are sent to the database owner as $\RLWE$ ciphertexts of degree $2^{12}$ encrypting the polynomial representation of the scoring functions, using the evaluation of arbitrary function technique we will review in Section~\ref{sections:large-domain}.
The database owner then homomorphically uses the repacking algorithm to repack the results in $\RLWE$ ciphertexts (steps (i) and (ii)).
This enables to densely pack up to $2^{12}$ scores in a single $\RLWE$ ciphertext.
Since an $\RLWE$ ciphertext of degree $2^{12}$ only allow for a modulus of up to $108$ bits to be secure, they are merged into $\RLWE$ ciphertexts of degree $2^{16}$ to enable bootstrapping to a larger modulus (e.g. $1540$ bits) enabling to continue performing further homomorphic evaluations.
This merging step also reduces the number of ciphertexts by a factor of $2^{16}/2^{12} = 16$, with each ciphertext encrypting up to $2^{16}$ scores.
Finally, the database owner bootstraps and homomorphically encodes theses large degree $\RLWE$ ciphertexts to enable encrypted SIMD arithmetic and leverage batching when evaluating the threshold functions on the encrypted scores (steps (iii), (iv) and (v)).

\paragraph{Relations to existing works on Private Function Evaluation (e.g., via universal circuits)} Private Function Evaluation (PFE) protocols enables two parties $P_A$ and $P_B$ owning respective private inputs $x_A$ and $x_B$ to learn $f(x_A,x_B)$ where only one of the two parties knows the function $f$. Several papers (\cite{STOC:Valiant76,FC:KolSch08,EC:KisSch16,EC:MohSad13,C:LYZZLH21,JC:AGKS20} for example) have focused on constructing PFE for arbitrary polynomial-size functions.
A well-known approach consists of relying on secure evaluations of universal circuits (UC), which takes a circuit $C$ of size $n$ from a sequence of circuits $\{\mathsf{C}_n\}_{n\in \mathbb{N}}$ and an input $x$ and outputs $\mathsf{C}_n(C,x)$ which corresponds to the evaluation of $C$ on $x$.
In that case, the PFE protocol inherits from the properties of the underlying secure function evaluation protocol used to evaluate $\mathsf{C}_n$ on private inputs $C$ and $x$. Valiant~\cite{STOC:Valiant76} proposed a size-optimal UC-based construction of size $\Theta(n\log n)$ that can simulate any Boolean circuit of size (less than or equal to) $n$.
Due to this optimal bound, as the circuit size increases, the complexity of UC-based PFE protocols becomes large as well.
Katz and Malka~\cite{AC:KatMal11} proposed a passively secure PFE protocol with linear complexity.
An implementation of the PFE protocol from~\cite{AC:KatMal11} proposed by Holz et al. in~\cite{ESORICS:HKRS20} shows that HE-based PFE outperforms UC-based PFE in communication for all circuit-size and in computation starting from circuit with a few thousand gates.
We make a brief comparison of the performance of our protocol with the one of~\cite{ESORICS:HKRS20} in Section~\ref{sec:uc_vs_fhe}.

\paragraph{Organization} Section~\ref{section:prelim} provides the necessary technical backgrounds to understand the homomorphic building blocks composing our use case. Section~\ref{section:model} introduces the system model for TETRIS and presents in details the homomorphic ring packing, ring merging and ring switching along with the HE-based PFE we use to handle computations over large datasets. Section~\ref{section:protocol} shows how to combine all the homomorphic ring operations to build TETRIS, our solution for private functional exploration of large datasets. Section~\ref{section:implem} details the performance of our construction. Finally, Section~\ref{section:conclusion} concludes the paper.

\section{Backgrounds}
\label{section:prelim}
\subsection{Notations}
We denote single elements (polynomials or numbers) in italics, e.g., $a$, and vectors of such elements in bold, e.g., $\bm{a}$. The $i$-th position of the vector $\bm{a}$ or the degree-$i$ coefficient of the polynomial $a$ will be denoted $a[i]$. The interval from the $i$-th to to $j$-th coefficients will be denoted as $[i:j]$. When the interval $[i:j]$ is a discretized subset of real values, we also denote it as $\mathbb{R}_{[i, j]}$.
We will use the operator $+$ for addition between polynomials, vectors or numbers, the operator $\cdot$ for point-wise multiplication between numbers, vectors or polynomials, $*$ for the polynomial convolution.
We denote $||\cdot||$ the infinity norm, $[\cdot]_{Q}$, $\lfloor\cdot\rfloor$, $\lfloor\cdot\rceil$ the reduction modulo $Q$, rounding to the previous and to the closest integer, respectively (coefficient-wise for polynomials), and $\langle \cdot, \cdot\rangle$ the dot product.
We denote by $\leftarrow\chi^{n}_{d}$ the act of sampling a vector (or polynomial) of size $n$ from the distribution $d$ (we omit $n$ when $n=1$).
Unless otherwise stated, logarithms are in base 2.

\subsection{Plaintexts and Ciphertexts Domain}\label{section:cyclotomicrings}
The $N$-th cyclotomic polynomial ring modulo $Q$ will be denoted $\mathcal{R}_{Q}^{N}[X] = \mathbb{Z}_{Q}[X]/(X^{N}+1)$ and defines the messages space and ciphertexts space.
Elements of $\mathcal{R}_{Q}^{N}[X]$ are all polynomials in $X$ with integer coefficients taken modulo $Q$ and of degree at most $N-1$.

We assume that the coefficients are centered, i.e. lie in $[-\lfloor Q/2\rfloor,$ $\lfloor Q/2\rfloor)$.
To keep notations concise, the variables $N$, $Q$ and/or $X$ will be omitted when denoting $\mathcal{R}$ when these do not need to be defined. 

The integer $5$ is a generator of order $N/2$ modulo $2N$, and along with the generator $-1$, it spans $\mathbb{Z}^{*}_{2N}$.
Therefore the set of automorphisms $\phi_i: X \mapsto X^{i}$ for $i \in (5^{j\in\mathbb{Z}_{N/2}}, -1)$ forms a group over $\mathsf{Gal}(K/\mathbb{Q})$ under composition, i.e. $\phi_{i}(a) + \phi_{i}(b) = \phi_{i}(a+b)$ and $\phi_{i}(a) \cdot \phi_{i}(b) = \phi_{i}(a \cdot b)$ for $a,b \in \fR$.\\

The ring $\ring{[X]}{N}{}$ contains multiple subrings with interesting properties, one of them being $\ring{[Y]}{N}{}$ for $Y = X^{2}$.
Since $\ring{[Y]}{N}{}\cong\ring{[X]}{N/2}{}$ we have $\ring{[X]}{N}{} \cong \ring{[X]}{N/2}{} + X * \ring{[X]}{N/2}{} \mod (X^{2} + 1)$.
This relation can be generalized to $Y = X^{N/n}$ for $n$ a power-of-two smaller than $N$ giving us the following isomorphism rule $\forall n|N$: 
$$\ring{}{N}{} \cong \overbrace{\ring{}{N/n}{} \otimes \dots \otimes \ring{}{N/n}{}}^{n}$$
The ring $\ring{[X]}{N}{}$ allows to encode polynomials in the ring (a.k.a in the \emph{coefficients}) or in the canonical embedding (a.k.a in the \emph{slots}). And messages can be homomorphically switched between these two encodings. And messages can be homomorphically switched between these two encodings and these follow different plaintext arithmetic. We detail in Appendix~\ref{appendix:plaintextdomains} the encoding in both plaintext domains.

\subsection{FHE Ciphertexts}
Let $s$ be polynomial in $\ring{[X]}{N}{}$ with coefficients sampled from a uniform distribution over $\{-1, 0, 1\}$ and with exactly $h$ non-zero coefficients. 
An RLWE ciphertext of a polynomial message $m$ under secret key $s$ is generated as $(c_0,c_1)=(-as + e+m, a)$ where $a$'s coefficients are uniform over $U(\mathbb{Z}_{Q})$ and $e$'s coefficients are sampled from a discrete Gaussian distribution with standard deviation $\sigma$. 
An RLWE ciphertext generated as such follows a distribution generated as $\mathcal{D}^{N, Q,h, \sigma}_{\textsf{RLWE}}$. Note that a fresh RLWE encryption of zero defines a polynomial in $s$ of degree $1$ that can be written as $\prlwe{}{s}{Q} = c_{0} + c_{1} * s = e$ with $||e||_{\infty}$ small.
After a few homomorphic operations, an RLWE encryption of zero defines a polynomial of degree $k$ that will be denoted $\prlwe{k}{s}{Q}$, if $k\neq 1$. And an (resp. the set of) $\RLWE$ encryption of $m$ of under secret key $s$ will be denoted $\RLWE_s(m)$ or $\RLWE(m)$ when the secret key doesn't need to be specified.

\subsection{Homomorphic Operations}
\label{section:homomorphic_operations}
\subsubsection{Keyswitching} keyswitching is at the core of the homomorphic properties of $\RLWE$-based encryption and used as a building block in a number of homomorphic operations most notably, homomorphic multiplication, morphism and ring splitting or ring merging. It makes use of an additional modulus $P$ which allows to control the noise growth.
\begin{itemize}
\item \textsf{SwitchKeyGen}($s$, $s'$$, \bm{w}$): 
For $s, s' \in \ring{}{N}{QP}$ and an integer decomposition basis of $\beta$ elements $\bm{w}=(w_{0},\cdots, w_{\beta-1})$, \textsf{Switch}-\textsf{KeyGen} returns the keyswitching key from $s$ to $s'$, $\textsf{swk}_{(s\rightarrow s')}=(\textsf{swk}^{(0)}_{(s\rightarrow s')},$ $\dots, \textsf{swk}^{(\beta-1)}_{(s\rightarrow s')})$, where $\textsf{swk}^{(i)}_{(s\rightarrow s')} = \prlwe{}{s'}{QP}+w_{i}Ps$;

\item 
$\textsf{SwitchKey}(d, \textsf{swk}_{s\rightarrow s'},\bm{w})$: decomposes $d \in \ring{}{N}{Q}$ in base $\bm{w}$ such that $d = \langle\bm{d}, \bm{w}\rangle$ and returns $\prlwe{}{s'}{Q} + ds = \lfloor P^{-1} \cdot \langle\bm{d}, \textsf{swk}_{s\rightarrow s'}\rangle\rceil \in (\ring{}{N}{Q})^{2}$ for $P^{-1} \in \mathbb{R}$.
\end{itemize}

\subsubsection{Advanced Homomorphic Operations for Scheme Switching} we now describe how the keyswitching operation $\textsf{SwitchKey}$ is used to perform some of the key homomorphic operations and give further details in Appendix~\ref{appendix:advanced_operations} on how these operations are processed using keyswitching.
\begin{itemize}
\item \textsf{Relinearization}: the multiplication between two $\RLWE$ ciphertexts $\prlwe{}{s}{}$ returns a new $\RLWE$ ciphertext $\prlwe{2}{s}{}$, and the degree will further increase if we perform additional multiplication with non-plaintext operands.\\

\item \textsf{Automorphism}: an automorphism $\phi$ applied on an $\RLWE$ ciphertext $\prlwe{}{s}{}$ maps it to a new RLWE ciphertext $\prlwe{}{\phi(s)}{}$.
The keyswitching operation enables to re-encrypt $\prlwe{}{\phi(s)}{}$ to $\prlwe{}{s}{}$ by evaluating $\textsf{SwitchKey}(\prlwe{[1]}{\phi(s)}{}, \textsf{swk}_{\phi(s)\rightarrow s}) + \prlwe{[0]}{\phi(s)}{}$;\\

\item \textsf{Ring Splitting}: it is possible to split an RLWE ciphertext $\prlwe{}{s}{} + m$ with $m,s\in\ring{}{N}{}$ into two ciphertexts of half the dimension $(\prlwe{}{s'}{} + m_{0}, \prlwe{}{s'}{} + m_{1})$ with $m_{0}, m_{1}, s'\in\ring{}{N/2}{}$ such that $m(X) = m_{0}(Y) + X*m_{1}(Y)$ for $Y = X^{2}$;\\

\item \textsf{Ring Merging}: it is possible to merge two RLWE ciphertexts $d_{0} = \prlwe{}{s'}{} + m_{0}$ and $d_{1} = \prlwe{}{s'}{} + m_{1}$ with $m_{0}, m_{1}, s'\in\ring{}{N/2}{}$ into an RLWE ciphertext of twice the dimension $\prlwe{}{s}{} + m$ with $m,s\in\ring{}{N}{}$ such that $m(X) = m_{0}(Y) + X*m_{1}(Y)$ for $Y = X^{2}$.\\

\end{itemize}

\subsection{Approximate Homomorphic Encryption}

The CKKS scheme by Cheon et al. \cite{CKKS2017} is an $\RLWE$ homomorphic-encryption scheme that enables SIMD arithmetic over $\mathbb{C}^{n}$.
Since its introduction, this scheme has grown in popularity, as it is currently the most efficient scheme for performing encrypted fixed-point arithmetic over complex numbers.

The setup of the scheme is done by instantiating a secure $\RLWE$ distribution $\mathcal{D}_{N, Q,h, \sigma}^\RLWE$, and a fresh encryption of $m$ is of the form $\prlwe{}{s}{Q} + m$.
This scheme is said to be approximate because the error is mixed with the message such that when we evaluate the decryption of $\RLWE_s(m)$ we obtain $m + e$ which is an approximation of $m$.

The scheme makes use of the canonical embedding to enable SIMD operations over $\mathbb{C}^{n}$ in the encrypted domain. We provide a full description in Appendix~\ref{appendix:plaintextdomains} on how the encoding works in the canonical embedding. The basic operations (addition, multiplication, rotation, conjugation) are sufficient to evaluate linear transformations and polynomial functions over $\mathbb{C}^{n}$.
Such linear transformations include the ability to homomorphically switch between the \emph{slots} and \emph{coefficients} encoding.

The encoding procedure uses fixed-point arithmetic with a scaling factor $\Delta$ to emulate $\mathbb{R}$ on $\mathbb{Z}$.
The consequence is that the multiplication of two messages scaled by $\Delta_{0}$ and $\Delta_{1}$ respectively returns a new message scaled by $\Delta_{2} = \Delta_{0}\Delta_{1}$.
It is easy to see that this scaling factor can have an exponential growth if not controlled, thus a \emph{rescaling} operation is required after each multiplication to ensure a linear growth.
This rescaling operation is straight-forward: truncate the lower bits of the ciphertext by evaluating $\round{1/\Delta}$ on the coefficients.
However, given a ciphertext $\prlwe{}{s}{Q}+m$, the rescaling procedure will return a new ciphertext in the ring $\prlwe{}{s}{Q/\Delta}+\round{m/\Delta}$ (we assume that $\Delta$ is a factor of $Q$).
After $i$ rescaling operations, we will get a ciphertext with a modulus $\Delta < Q/\Delta^{i} < \Delta^{2}$ and no further \emph{rescaling} operation can be carried out, therefore no more multiplication.
Such ciphertext is said to be \emph{exhausted} and then there are two options available: decrypt or homomorphically re-encrypt the ciphertext to a modulus $Q' > \Delta^{2}$ such that at least one additional rescaling (i.e. multiplication) can be evaluated before it becomes exhausted again. This homomorphic re-encryption is called \emph{bootstrapping} and it enables the evaluation of circuits of arbitrary depth under encryption. 

\subsection{Bootstrapping for Approximate Homomorphic Encryption}\label{section:bootstrapping}

For an RLWE distribution $\mathcal{D}^{N, Q,h, \sigma}_{\textsf{RLWE}}$, $X = Y^{N/n}$, a polynomial message $m\in\ring{}{n}{}$ and $\Delta$ a scaling factor, the bootstrapping procedure of the CKKS scheme~\cite{EC:CHKKS18} takes as input an \emph{exhausted} ciphertext $\prlwe{}{s}{q}+m(Y)$ with $\Delta < q < \Delta^{2}$ and returns a new ciphertext $\prlwe{}{s}{Q'}+m(Y)$ for $\Delta^{2} < Q' < Q$, enabling further homomorphic evaluations.
The procedure consists of the following four steps: \textsf{ModRaise}, \textsf{CoeffsToSlots}, \textsf{EvalMod}, and \textsf{SlotsToCoeffs} which are recalled in Appendix~\ref{appendix:bootstrapping}.

\section{System Model}
\label{section:model}
\subsection{Use case Context}
\label{subsection:context}
Our setting is the following: a scientist would like to conduct a horizontal study requiring patients’ datasets by applying a very specific combination of attributes, which the scientist would like to keep private.
We consider a two-party protocol between the database owner holding the patients' data and the scientist, who holds a private function of the form defined at the end of this subsection.
To be funded, the scientist would like to assess if there are enough patients meeting the selected research criterion.
During a first phase called \emph{homomorphic private function evaluation}, the database owner applies the scientist's encrypted criteria over the patient’s data.
This phase computes an encrypted result which is the evaluation of the encrypted attribute functions over the patient’s dataset.
The outcome of this homomorphic computation provides a preliminary validation over the patients’ datasets by computing a score per patient with respect to some metrics such as the age, the blood pressure or the blood sugar level.
Then in a second phase called \emph{homomorphic thresholds evaluation}, the database owner evaluates two consecutive threshold functions, but privately parameterized by the scientist, over an encrypted batch of patients’ data built from the output of the first phase.
Overall, the computation boils down to answering the following question: 
\begin{quote}
\emph{Given a database of $p$ patients and a set of $h$ criteria, are there at least $p'\leq p$ patients meeting at least $h'\leq h$ criteria?}
\end{quote}

We will show how this question can be answered efficiently for a database of hundreds of thousands of patients with dozens of attributes while still ensuring privacy with respect to the database owner and the scientist. 


\subsection{Threat Model}
\label{sec:threat_model}
Both the database owner and scientist are assumed to be \emph{honest-but-curious}, meaning that both follow the protocol honestly but still try to infer as much as possible from the each other's private inputs (the database and the private functions respectively). To model this setting, we consider the three following security properties: 
\begin{enumerate}[label=\arabic*)]
\item
Privacy with respect to the database owner: this property captures the fact that the database owner does not learn the response output to the question. 
\item
Privacy with respect to the scientist: this property captures the fact that the scientist does learn more information on the patients’ datasets than what can be inferred from the database owner output. 
\item 
Bounded extractable information: since the scientist has full control over the question asked to the database we need a way to quantify the amount of information extracted from the database by the scientist per request after decryption. 
Assuming the database owner evaluates the function honestly, this notion aims to quantify the amount of information that is sent back to the scientist, even in the case where the latter sends a malicious function. In our case, the extractable information is defined as the domain size of the decrypted response i.e. number of bits sent back by the database owner. Hence in our framework, the extractable information is bounded by 1 bit per query.
\end{enumerate}

\emph{Potential leak regarding the selected attributes}: in order to achieve a concretely-efficient solution for multivariate functions, the scientist uploads an attributes-selection plaintext matrix which is used to select and/or concatenate sets of attributes together. In the case where this matrix contains exactly the attributes that will be used, the database owner learns which attributes are being grouped, but not the combining function.
However, the scientist can include dummy attributes and build the private function to not take these dummy attributes into account.
In that case, the database owner learns the sets of attributes that are used but not necessarily which of them exactly.
In particular, if the scientist selects all the attributes, there is no leakage regarding the attributes used for univariate functions.
Quantifying how much information is leaked will highly depend on the attributes-selection matrix being sent by the scientist. As the database owner will not learn the final output at all, property (1) is expressed as an all-or-nothing privacy property. This means that in our framework, we consider the database owner can learn which attributes are being grouped or selected, but not the combining function and its output.\\

\emph{Potential leak by using a malicious function}: as discussed above, the scientist might attempt to send a malicious function in order to learn more information on the patients dataset than what the database owner would allow.
In our use case, the database owner processes the patients’ data in two steps. In the first phase (homomorphic private function evaluation), the patients’ data are processed in the cleartext domain by applying the encrypted function sent by the scientist and the output is encrypted. 
Considering the form of the statement tested by the scientist (as expressed at the end of subsection~\ref{subsection:context}), the scientist explicitly asks for a threshold computation on a private (scoring) function.
Regardless of the parameters of the thresholds evaluated by the database owner, the amount of information is bounded to $1$ bit per request (the output domain size of the threshold function).
If the scientist acts dishonestly by sending a malicious encrypted function in the first phase, we argue that the database owner still has full control on how much information is returned to the scientist for each request.\\

\emph{Dealing with multiple malicious requests:} In our framework, we can not avoid attacks from a scientist who would use high dimensional requests to extract information that it should not, for example by using multiple requests to try to infer if a specific individual is part of the database.
Such attack becomes unavoidable as soon as any amount of information about the content of the database is returned to the scientist. In that case, the amount of information extracted by the scientist relates to the number of requests that would be needed to achieve the desired outcome.
However, limiting the throughput of information that is returned by request to the scientist, or bounding the number of request that a scientist can make, in combination with differential privacy techniques can help mitigate this kind of attacks. 

We now give further details on the protocol interactions between the scientist and the patients database owner.

\subsection{TETRIS High-Level Overview}
The protocol is divided in two phases, a \emph{homomorphic private function evaluation phase} and a \emph{homomorphic thresholds evaluation phase} which work as follow: 
\begin{enumerate}
\item \emph{Homomorphic private function evaluation phase.} The inputs to the homomorphic private function evaluation phase are:
\begin{itemize}
\item Scientist: an attributes-selection plaintext matrix $M$ of size $h\times m$ and a list of $m$ encrypted scoring functions over the features represented as a vectors of $\RLWE$ encryptions. 
\item Database owner: a database $P$ represented as a $p\times h$ matrix, where $p$ is the number of patients and $h$ is the number of features.
We will denote by $P[i]$ the $i$-th entry of the database (i.e. the matrix row) by $P[i][j]$ and its $j$-th feature. 
\end{itemize}
Throughout this phase, the goal of the attributes-selection plaintext matrix is to enable features selection, but it can also be used to produce a linear combination of features.
The scientist starts by sending the encrypted functions together with the attributes-selection plaintext matrix to the database owner.
At the end of the homomorphic private function evaluation phase, the database owner obtains the encrypted evaluation of the functions applied to the database.\\
\item \emph{Homomorphic thresholds evaluation phase.} The database owner homomorphically combines the patients' scoring results and homomorphically evaluates a first threshold (local threshold), which corresponds to answering the question \emph{does $P[i]$ meets $h'\leq h$ criteria} (for all $i$)?
The database owner then aggregates the results and homomorphically evaluates a second threshold (global threshold), which corresponds to answering the question \emph{are there at least $p'\leq p$ patients fulfilling the first question?}
\end{enumerate}

\section{Building blocks}
In this section, we introduce the different building blocks that are required by our construction. 

\subsection{Ring Repacking}
\label{section:repacking}
Several (re)packing algorithms were proposed in the literature \cite{AC:CGGI17,ACNS:CDKS21,SP:LHHMQ21,JMC:BGGJ20,EPRINT:KDECLG21,C:BCKPS23}, some of them are proposed to pack a list of $n$ encryptions of scalar messages $x_i$ (LWE encryptions) to an $\RLWE$ encryption of $\sum_{i=0}^{n-1}x_i\cdot X^i$, and the procedure is called \emph{ring packing} in that case.
Ring repacking refers to the operation that maps a list of $\RLWE$ encryptions of some polynomial messages to an $\RLWE$ encryption whose message coefficients are taken from the input list of messages coefficients.
In our work, we will rely on the iterative version of the \emph{ring repacking} algorithm proposed in~\cite{ACNS:CDKS21,EPRINT:KDECLG21}, that we recall in Algorithm~\ref{alg:repacking}.
This iterative version was recently used in~\cite{FHEorg:BosIza24} in the context of homomorphic evaluation of (public) arbitrary functions over encrypted inputs. 
The goal of the \emph{ring repacking} algorithm is to repack $n$ $\RLWE$ ciphertexts encrypting each a polynomial with $x_i$ in its constant coefficient for $i\in[0,n-1]$ into an $\RLWE$ ciphertext encrypting $\sum_{i=0}^{n-1}x_i\cdot X^i$.

For sake of simplicity, we take the number of input ciphertexts to be equal to $N$ i.e the degree of the polynomial ring, but the algorithm can be easily generalized to any number of inputs less than $N$ just by taking zero values for the remaining coefficients and if $n>N$ we can split the $n$ ciphertexts into batches of up to $N$ ciphertexts and evaluate a repacking per batch. Regarding its complexity, Algorithm~\ref{alg:repacking} requires $N$ homomorphic automorphisms evaluations and $\log N$ distinct switching keys. 

\begin{algorithm}
\caption{\textsc{Ring Repacking} of $N$ $\RLWE$ encryptions into one $\RLWE$}
\label{alg:repacking}
\begin{algorithmic}
\State \underline{Inputs:} $N$ $\RLWE$ ciphertexts 
$c_1 = (a_0,b_0), \dots, c_n = (a_{N-1},b_{N-1})$ where $(a_i,b_i)$ is an encryption under secret key $s$ of a polynomial $m_i(X)$ whose constant coefficient is $m_i[0]$; the keyswitching keys $\textsf{swk}_{\phi_g(s)\rightarrow s}$ for $g=5^{2^{i-1}}$ and $i\in [0,\log N-1]$ for homomorphically evaluating the automorphisms $\phi_g$.
\State \underline{Output:} an $\RLWE$ encryption of $m'(X) = \sum_{i=0}^{N-1} m_i[0]\cdot X^i$.\\

\State \For{$i\gets0 < N$}{
$c_{i} \leftarrow N^{-1}c_{i}$~~// \small{pre-multiplication by $1/N\mod Q$}\\
}
\State \For{$i\gets0 < \log N$}{
\State $t= N/2^{i+1}$ ~~// \small{number of remaining steps }
\State \uIf{$i = 0$}{
$g\leftarrow 2N-1$
}\Else{
$g\leftarrow 5^{2^{i-1}} \mod 2N$
}
\State \For{$j \gets 0 < t$}{ 
\State $c_j \leftarrow c_j + c_{j+t} \cdot X^{t}+\phi_g(c_j - c_{j+t} \cdot X^{t})$
}
}
\State return $c_0$
\end{algorithmic}
\end{algorithm}

\begin{lemma}\label{lem:repackingcorrectness}
Let $c_1 = (a_1,b_1), \dots, c_n = (a_N,b_N)$ be encryption of $m_1(X), \cdots, m_N(X)$. Algorithm~\ref{alg:repacking} takes as inputs the $c_i$ and returns an encryption of $\mu(X)=\sum_i \mu_i\cdot X^i$ such that $\mu_i=m_i[0]$.
\end{lemma}

We refer to~\cite[proof of Algorithm 2]{ACNS:CDKS21} for the proof of Lemma~\ref{lem:repackingcorrectness}. Regarding the noise growth induced by Algorithm~\ref{alg:repacking}, \cite{ACNS:CDKS21} does not provide an analysis, however its behavior is predictable and can easily be explained:
we can notice that at each of the $\log N$ steps of the algorithm, pairs of ciphertexts have their even coefficients that double and their odds coefficients that vanish before being added together.
Additionally, one of the ciphertexts goes through a keyswitching operation.
Therefore, given $e_{\textsf{ct}}$, the noise output of the previous step and $e_{\textsf{ks}}$ the noise of the keyswitching operation, the new noise after each iteration becomes $2e_{\textsf{ct}} + e_{\textsf{ks}}$. Over $\log N$ steps, this gives us an estimated noise growth of $\approx N(e_{\textsf{ct}} + e_{\textsf{ks}})$. Now since the initial ciphertext coefficients, and thus the noise, have been pre-scaled by $N^{-1}$, the final noise is actually $\approx e_{\textsf{ct}} + Ne_{\textsf{ks}}$. The noise growth is therefore independent of the initial noise and depends solely on the additive noise of the keyswitching operation, which can be parameterized through the gadget decomposition to achieve the desired bounds.

\subsection{Ring Merging and Schemes Switching}
\label{sections:switching}
Ring merging is employed in \emph{scheme-switching} which makes use of the CKKS bootstrapping which requires a large ring degree. We first explain the principle behind \emph{scheme-switching} and then where ring merging 
comes into play. To enable an efficient \emph{ring repacking} procedure, the functions that will be evaluated are encrypted in a ring of small degree $n$ and with a small modulus $Q_{0}$: $\ring{}{n}{Q_{0}}$. 
So the output of the ring repacking is in the coefficients domain and in the ring $\ring{}{n}{Q_{0}}$.
To enable SIMD arithmetic and further homomorphic operations, we need to homomorphically switch the values from the coefficients’ domain to the slots domain and we need to increase the size of the modulus. For the sake of completeness, the switching operations in the plaintext domains are detailed in Appendix~\ref{appendix:plaintextdomains}.
This can be done using the CKKS bootstrapping, however to be able to evaluate the bootstrapping circuit, with a large enough residual homomorphic capacity after its evaluation, we must increase the modulus from $Q_{0}$ to $Q_{\ell}$, and as a consequence the ring degree from $n$ to $N$ to maintain the same security. 
To switch from a ring degree of $n$ to $N$, we rely on \emph{ring merging}. \\

The goal of the \emph{ring merging} procedure is to pack a batch of $N/n$ ciphertexts in $\ring{}{n}{Q}$, each encrypting $\sum_{j=0}^{n-1} x_{i,j}X^j$ for $i\in[0,N/n]$ in a single ciphertext in $\ring{}{N}{Q}$ encrypting 
$\sum_{i=0}^{N/n-1}\left(\sum_{j=0}^{n-1} x_{i,j} \cdot X^{N/n}\right)\cdot X^i$. In other words, we can merge a list of $N/n$ ciphertexts of degree $n$ into a single $\RLWE$ ciphertext of degree $N$. Concrete values for \emph{scheme switching} operations will be given in Section~\ref{section:implem}.

\subsection{Large-Domain Private Function}
\label{sections:large-domain}
The goal of this building block is to evaluate an encrypted arbitrary function $f: \mathbb{R}_{[a, b]} \rightarrow \mathbb{R}_{[c, d]}$ on one or several plaintext values 
in $[a, b]$. 
There are several methods in the literature that show how to evaluate arbitrary functions.
However, as discussed in Section~\ref{section:intro} most of them are constrained in the number of inputs to keep the evaluation method sufficiently efficient in practice. 
Our method is based on an adaptation of a solution proposed by~\cite{PoPETS:IIMP22,FHEorg:BosIza24}, which outperforms known homomorphic LUT evaluation techniques
for domains as large as $2^{14}$ and higher while providing sufficient precision. While this method is not composable (i.e. the method cannot be recursively invoked), it still sufficient for our purpose) and it supports large domain size in an efficient manner; by way of illustration, \cite{FHEorg:BosIza24} reports a two-to-three orders of magnitude
improvements compared to previous evaluation methods. 
In their case, the inputs of the large domain function are encrypted and the function to be evaluated is encoded in the plaintext domain. We consider the dual scenario where the functions to be evaluated are encrypted and the inputs are in the plaintext domain. We review below the interactions between the scientist and the database owner, where the latter evaluates $f:\dom_f \mapsto \img_f$ (defined in the encrypted domain) on a set of 
points in $I\subseteq \dom_f$ (in the plaintext domain):
\begin{enumerate}
\item
Each point $i\in I:=\{i_0, i_1, \cdots,i_{I-1}\}$, where $I$ is held by the database owner in our case, is encoded in the exponent as $X^i$;\\
\item 
The scientist defines a polynomial representation of the function $f$ as $\vec{u}_f:=f(0)-f(N-1)\cdot X-f(N-2)\cdot X^2 - \cdots -f(1)\cdot X^{N-1}$ that is encrypted as an $\RLWE$ ciphertext, denoted $\vec{U}_f \in \langle \prlwe{}{s}{Q}+ \vec{u}_f \rangle$. 

For each element encoded as $X^i$ by the database owner, it holds that:
$$X^{i} \cdot \vec{u}_f = X^{i} \cdot \vec{U}_f \in \langle \prlwe{}{s}{Q}+(f(i)\cdot X^0 + \star)\rangle$$
which encrypts a polynomial whose constant coefficient is equal to $f(i)$. 
Using the ring-repacking procedure, the database owner maps the $\RLWE$ encryptions of 
$f(i_0)\cdot X^0 + \star$, $f(i_1)\cdot X^0 + \star$, $\cdots$ and $f(i_{I-1})\cdot X^0 + \star$ 
to an $\RLWE$ encryption $\vec{U}_f$ of $f(i_0)+f(i_1)\cdot X+f(i_2)\cdot X^2 + \cdots +f(i_{I-1})\cdot X^{I-1}$;\\

\item
The scientist receives and decrypts $\vec{U}_f$ to obtain $f(i_0)+f(i_1)\cdot X+f(i_2)\cdot X^2 +\cdots +f(i_{I-1})\cdot X^{I-1}$. He parses the result and obtains the consecutive evaluations on the target points in the decrypted polynomial coefficients. 
\end{enumerate}

\paragraph{Complexity} Using this methodology, the database owner homomorphically evaluates $2|I|$ plaintext-ciphertext products and $|I|$ automorphisms. At the second step, the database owner can alternatively send $|I|$ $\RLWE$ encryptions of $f(i)\cdot X^0 + \star$ for each $i\in I$ and the scientist would decrypt each of them independently to retrieve $f(i)$ for each $i\in I$. However, this method has two drawbacks. First, it will not permit further computations on the encryption of $f(i_0)+f(i_1)\cdot X+f(i_2)\cdot X^2 +\cdots +f(i_{I-1})\cdot X^{I-1}$ without using repacking. Second, it would incur an additional $|I|$ factor in the communication complexity from the database owner to the scientist side. Instead, using ring repacking enables the database owner to send back only one $\RLWE$ ciphertext, this is why we implement the first method. 

\paragraph{Security properties} In our use case, the points are processed in clear by the database owner who applies the encrypted function sent by the scientist, during the homomorphic private function evaluation phase.
As discussed in Subsection~\ref{sec:threat_model}, while the database owner knows which of the attributes are used, it does not learn the function itself (ie. how the attributes are grouped) during this phase since both the function and its output are encrypted.
Then, in the homomorphic thresholds evaluation phase, the database owner applies two thresholds privately parametrized by the scientist over an encrypted selected dataset. 
The privacy property with respect to the scientist is guaranteed if the underlying $\RLWE$ encryption scheme is semantically secure which holds under the RLWE assumption.
We provide concrete parameters for our use case in Section~\ref{section:implem}. In the homomorphic thresholds evaluation phase, the database owner controls the information that is computed over the encrypted selected patients' dataset.
Assuming the RLWE encryption scheme is correct, the scientist does not learn more than what the evaluation of the function over a selected patient’s dataset provides; this means that after decryption the scientist does not learn more that $1$ bit of information per request. As defined by the bounded extractable information property, the amount of information is quantified per request as we cannot avoid attacks where the scientist tries to infer more than $1$ bit by combining evaluation results of many malicious functions requests. Hence, the privacy property with respect to the database owner and the bounded extractable information per request property are both guaranteed.

\section{TETRIS Specifications}
\label{section:protocol}

We first give the concrete setting for the private database exploration use case and detail how the core functions are defined and implemented in the encrypted domain. We then present the circuit's structure in TETRIS and its different steps.

\subsection{Dataset, Attributes Selection \& Criteria}
\label{section:practicalcase_context}

\subsubsection{Dataset}
We generate a synthetic dataset of $p=2^{19}$ entries and $h=16$ features.
Features are drawn from a Gaussian distribution of standard deviation $\sigma=1$ and mean $\mu = 1$ and are bounded to the interval $[0, 2)$.

\subsubsection{Attributes Selection \& Criteria}
The scientist is assumed to provide an attributes-selection plaintext $M$ matrix $h\times m$.
The primary goal of this matrix is to enable features selection, but it can also be used to produce a linear combination of attributes before evaluating a scoring function.
In our practical case, we use $16$ scoring functions of the form:
\[
f: \mathbb{R}_{[a, b)}\rightarrow\mathbb{Z}_{10},
\]
where we take $a=0$ and $b=2$.

\subsection{Scheme Switching Operations}

\subsubsection{Ring Repacking} Each evaluation of the private function $\RLWE(\bm{u}_{f_{j}})$, for $j\in [0,m-1]$ on a value $x$ returns an $\RLWE$ ciphertext encrypting a polynomial whose constant coefficient is the value $f_j(x)$.
We recall that the goal of this building block is to repack $n$ $\RLWE$ ciphertexts, for $i\in[0,n-1]$ encrypting $\RLWE(\bm{u}_{f_{i}})$ in their constant coefficient into a single $\RLWE$ ciphertext encrypting $\sum_{i=0}^{n-1} f_j(x_{i})\cdot X^{i}$, producing a densely packed ciphertext in the ring $\ring{}{n}{Q_{0}}$ for some possibly distinct $j\in[0,m-1]$. In our case, the $f_j$'s are the final scoring functions for each row.
If $p<n$ (we recall that $p$ is the number of rows in the database), we can just take zero values for the remaining coefficients.
This homomorphic operation is done by using the repacking technique described in Algorithm~\ref{alg:repacking}.
\[
(\RLWE(\textsf{score}_{P[0]}), \dots, \RLWE(\textsf{score}_{P[n-1]})) \in \ring{}{n}{Q_{0}}
\]
\[
\Bigg\downarrow \textsf{Repack}
\]
\[
\RLWE\left(\sum_{i=0}^{n-1}\textsf{score}_{P[i]}X^{i}\right) \in \ring{}{n}{Q_{0}}
\]

\subsubsection{Ring Merging}

The output of the \emph{ring packing} is in the coefficients domain and in the ring $\ring{}{n}{Q_{0}}$, with a small modulus $Q_0$ to enable more efficient homomorphic operations. However, in order to support additional homomorphic operations and SIMD encoded arithmetic, we make use of the CKKS bootstrapping, which requires to increase the modulus and ring degree to maintain the same level of security, as explained in Section~\ref{sections:switching}.
Given a batch of $N/n$ ciphertexts, each encrypting $\sum_{j=0}^{n-1}f(x_{i, j})$ for $0\leq i < N/n$, the \emph{ring merging} operation merges them together into a single ciphertext encrypting $\sum_{i=0}^{N/n-1}\left(\sum_{j=0}^{n-1}f(x_{i, j})\cdot X^{N/n}\right)\cdot X^{i} \in \ring{}{N}{Q_{0}}$.
If a batch contains less than $N/n$ ciphertexts, it is padded with noiseless zero ciphertexts.
\[
\begin{array}{c}
\left(\RLWE\left(\sum\limits_{i=0}^{n-1}\textsf{score}_{P[i]}\cdot X^{i}\right), \dots, \RLWE\left(\sum\limits_{i=0}^{n-1}\textsf{score}_{P[i+n]}\cdot X^{i}\right)\right)\\
\in (\ring{}{n}{Q_{0}})^{N/n}
\end{array}
\]
\[
\Bigg \downarrow \textsf{Merge}
\]
\[
\begin{array}{c}
\RLWE\left(\sum\limits_{i=0}^{N/n-1}\left(\sum\limits_{j=0}^{n-1}\textsf{score}_{P[j+in]}\cdot X^{N/n}\right)\cdot X^{i}\right)\\
\in \ring{}{N}{Q_{0}}
\end{array}
\]

\subsection{Bootstrapping}
\label{sec:practicalcase:implementation}

Given a ciphertext in $\ring{}{N}{Q_{0}}$, the primary goal of the CKKS bootstrapping is to produce a new ciphertext in $\ring{}{N}{Q_{\ell}}$, with $Q_{\ell}\gg Q_{0}$ encrypting the same message. Appendix~\ref{appendix:bootstrapping} provides a full description of the CKKS bootstrapping. We review here the main steps. The CKKS bootstrapping is divided into four steps:
\begin{enumerate}
\item \texttt{ModRaise}: raise the modulus from $Q_{0}$ to $Q_{L}$.
\item \texttt{CoeffsToSlots}: homomorphically encode the underlying message.
\item \texttt{EvalMod}: homomorphically evaluate a modular reduction by $Q_{0}$.
\item \texttt{SlotsToCoeffs}: homomorphically decode the underlying message.
\end{enumerate}
The output ciphertext is at level $Q_{\ell} = Q_{L-k}$, where $k$ is the depth of the bootstrapping circuit.
The default bootstrapping aims to keep the message in the same encoding domain, making it independent of the underlying message encoding.
However, in addition to raising the modulus from $Q_{0}$ to $Q_{\ell}$, one of our goals is to switch the messages from the coefficients domain to the slots domain. We observe that this can be obtained for free by skipping step 4 of the bootstrapping circuit.
We denote such a bootstrapping circuit as \textsf{Half-BTS}.
\[
\RLWE\left(\sum\limits_{i=0}^{N/n-1}\left(\sum\limits_{j=0}^{n-1}\textsf{score}_{P[j+in]}\cdot X^{N/n}\right)\cdot X^{i}\right) \in \ring{}{N}{Q_{0}}
\]
\[
\Bigg \downarrow \textsf{Half-BTS}
\]
\[
\begin{array}{c}
\RLWE\left(\mathsf{Ecd}\left(\sum\limits_{i=0}^{N/n-1}\left(\sum\limits_{j=0}^{n-1}\textsf{score}_{P[j+in]}\cdot X^{N/n}\right)\cdot X^{i}\right)\right)\\
\in \ring{}{N}{Q_{\ell}}
\end{array}
\]
Note that, although not illustrated here for sake of readability, \textsf{Half-BTS} actually returns two ciphertexts, each encrypting a vector of $N/2$ encoded scores.
The reason is that during the \texttt{CoeffsToSlots} step, the $N$ scores are encoded as an $N/2$ complex vector, with half of the the scores in the real component and the other half in the imaginary component. And since the step function is defined on the reals, we need to extract this imaginary part into a real vector.

\subsection{Core Functions Implementation}
\label{section:circuit}

\subsubsection{Encrypted Scoring Functions}\label{subsubsec:encrypted_scoring_function}
The private scoring function is evaluated for the $h$ attributes of each of the $p$ entries of the matrix $P$. Each of the attribute scoring function $f_j: \mathbb{R}_{[a, b]} \rightarrow \mathbb{Z}_p$ is encoded as a polynomial $\bm{u}_{f_{j}}\in\ring{}{n}{}$. We define a discretization factor of $\frac{1}{N}$ and the function $f_j$ is encoded as:

\[
\bm{u}_{f_{j}} = f\left(0\right) - \sum_{i=1}^{N-1} f_j\left(g^{-1}\left(\tfrac{N-i}{N}\right)\right)\cdot X^{i},
\]
where $g: \mathbb{R}_{[a, b]}\rightarrow\mathbb{R}_{[0, 1]} $ and $g(x) = \frac{1}{2}\left(\frac{2x-b-a}{b-a}+1\right)$. 

Then for $x\in[a, b]$, we have
\[
X^{\lfloor Ng(x)\rceil} \cdot \RLWE(\bm{u}_{f_{j}}) \approx \RLWE(f_j(x)\cdot X^0 + \star)
\]

with an error bounded by $|f_j(x) - f_j(x\pm \frac{b-a}{N})|$. We stress that if $|x_{i}-x_{j}| \geq (b-a)/N, \forall x_{i},x_{j}$ then this error does appear since all possible input points can exhaustively be covered.
If $|x_{i}-x_{j}| \leq (b-a)/N$ (assuming that $N$ cannot be further increased), then the error depends on the smoothness of the function and in such cases it can be mitigated by ensuring that $f$ is an $\mathcal{L}$\emph{-Lipschitz} function over the considered interval\footnote{$f$ is an $\mathcal{L}$\emph{-Lipschitz} function if $\forall x_{i}, y_{i}$ we have $|f(x_{i})-f(x_{j})| \leq \mathcal{L}|x_{i}-x_{j}|$.}.

In our use case, the functions are encrypted by the scientist and sent to the database owner for evaluation over the features, each function representing a private selection criterion.
The results of each scoring function are aggregated together to form a final score for each of the $p$ entries of the database. This final score for row $i$ is denoted $\textsf{score}_{P[i]}$ and defined as:
\[
\RLWE(\textsf{score}_{P[i]}) = \sum_{j=0}^{15} \RLWE(\bm{u}_{f_{j}})\cdot X^{P[i][j]},
\]

\subsubsection{Private Threshold Functions}
Let the threshold function parameterized by $t$ be:
\[
\texttt{thresh}_{t}(x):
\begin{cases}
1 & \text{ if } x > t\\
0 & \text{otherwise}.
\end{cases}
\]
\noindent Note that $\texttt{thresh}_{t}(x)$ is equivalent to $\texttt{step}(x - t)$, where
\[
\texttt{step}(x):
\begin{cases}
1 & \text{ if } x > 0\\
0.5 & \text{ if } x = 0\\
0 & \text{otherwise}.\\
\end{cases}
\]

\noindent Assuming that $|x_{i}-x_{j}| \geq \epsilon$ for some $\epsilon$, and that $|x - t| = k\epsilon$ for some integer $k$, we can instead evaluate $\texttt{step}(x - t + \epsilon/2)$, which will prevent the case $x - t = 0$.\\

For the approach we use, evaluating the threshold functions requires to scale $x-t$ to the interval $[-1,1]$.
For the first local threshold, this is done by normalizing with the factor $\frac{1}{\sum_{j=0}^{15} \max F_j}$ and for the second global threshold by using a normalization factor of $\frac{1}{p}$.\\

Furthermore, we can make $\texttt{thresh}_{t}(x)$ private by providing $\RLWE(t)$ to the evaluator, computing $\texttt{step}(\RLWE(x) - \RLWE(t))$:\\
\medskip

\[
\begin{array}{c}
\tiny{\RLWE\left(\mathsf{Ecd}\left(\sum\limits_{i=0}^{N/n-1}\left(\sum\limits_{j=0}^{n-1}\textsf{score}_{P[j+in]}\cdot X^{N/n}\right)\cdot X^{i}\right)\right)} \\
\in \ring{}{N}{Q_{\ell}}
\end{array}
\]

\[
\Bigg\downarrow\texttt{thresh}_{t}
\]

\[
\begin{array}{c}
\RLWE\left(\mathsf{Ecd}\left(\sum\limits_{i=0}^{N/n-1}\left(\sum\limits_{j=0}^{n-1}\texttt{thresh}_{t}(\textsf{score}_{P[j+in]})\cdot X^{N/n}\right)\cdot X^{i}\right)\right)\\
\in \ring{}{N}{Q_{\ell}}
\end{array}
\]
Evaluating $\texttt{thresh}_{t}$ with approximate homomorphic encryption is a challenging task.
We chose the approach of~\cite{LLNKSign2020}, in which they approximate the sign and step functions in the interval $[-1, -2^{-\alpha}]\cup[2^{-\alpha}, 1]$ as a composition of low-degree minimax approximations.
Since their approach requires the input to be in the interval $[-1, -2^{-\alpha}]\cup[2^{-\alpha}, 1]$, the input values need to be normalized to this interval.
Because we are evaluating thresholds over discrete values, we can ensure that the threshold always returns the correct value by setting $2^{-\alpha} < 1/p$, where $1/p$ is the normalization factor.

\subsection{Putting Altogether}

\subsubsection{Illustration of the Exploration database procedure}
\label{section:illustration}

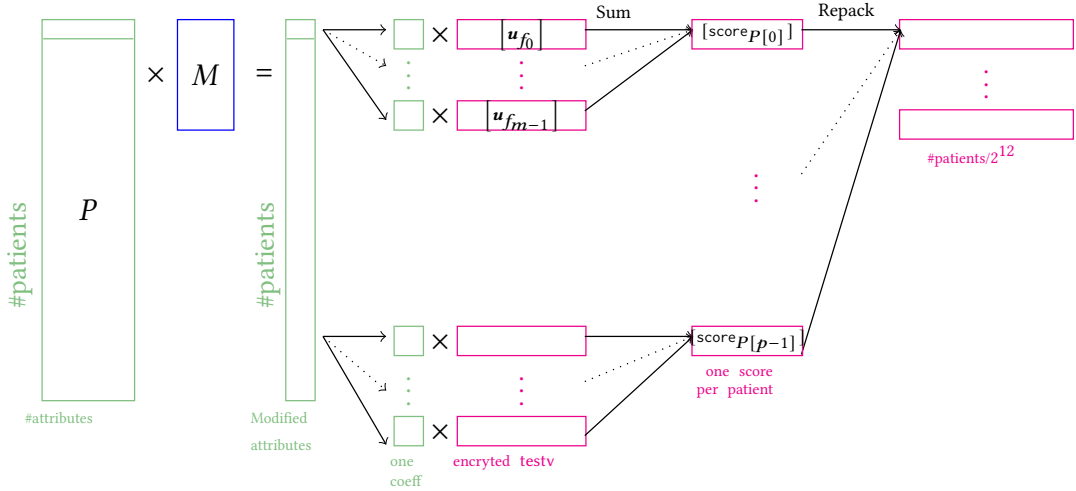
\begin{figure*}
\centering
\scalebox{1.2}{
\begin{tikzpicture}[decoration={brace}][scale=0.5]
\tikzstyle{operation}=[->,>=latex]

\draw (-12,12) node (P0) {}; 
\draw (-11.2,8) node (P00) {}; 
\path [draw, -] (-12.1,11.9) [color=green!50!black!50] -- (-11.1,11.9) {} ;
\draw[color=green!50!black!50] (P0.north west) rectangle (P00.south east) {};
\node at (-12.5,9.5) [color=green!50!black!50,text width = 0.1pt] {\rotatebox{90}{\#patients}};
\draw (-11.6,10) node {\large{$P$}};

  \draw (-10.85,11.5) node {$\times$};      
\draw (-10.5,12) node (L0) {}; 
\draw (-10.1,11) node (L00) {}; 
\draw[color=blue] (L0.north west) rectangle (L00.south east) {};
\draw (-10.3,11.5) node {\large{$M$}};

\draw (-9.65,11.5) node {$=$};

\node at (-9.75,9.5) [color=green!50!black!50,text width = 0.1pt] {\rotatebox{90}{\#patients}};
\draw (-9.3,12) node (P0) {}; 
\draw (-9.2,8) node (P00) {}; 
\draw[color=green!50!black!50] (P0.north west) rectangle (P00.south east) {};
\path [draw, -] (-9.4,11.9) [color=green!50!black!50] -- (-9.1,11.9) {} ;

\node at (-7,13) {\textsc{Operations on the Server's side - Homomorphic private function evaluation phase}};
\path [draw, ->] (-9,12) -- (-8.3,12) {} ;
\path [draw, ->,dotted] (-9,12) -- (-8.3,11.6) {} ;
\path [draw, ->] (-9,12) -- (-8.3,11) {} ;

\draw (-8.1,12) node (c0) {}; 
\draw (-8,11.9) node (c00) {}; 
\draw [color=green!50!black!50] (c0.north west) rectangle (c00.south east);
\path (-8.05,11.6)  -- node[color=green!50!black!50]{\vdots}  (-8.05,11.6);
\draw (-8.1,11.1) node (c0) {}; 
\draw (-8,11) node (c00) {}; 
\draw [color=green!50!black!50] (c0.north west) rectangle (c00.south east);

\draw (-7.7,11.95) node {$\times$};
\draw (-7.7,11.05) node {$\times$};

\draw (-7.4,12) node (c0) {}; 
\draw (-6.2,11.9) node (c00) {}; 
\draw [color=magenta] (c0.north west) rectangle (c00.south east);
\draw (-6.8,11.95) node {\tiny{$\left[\bm{u}_{f_{0}}\right]$}};

\path (-6.7,11.6)  -- node[color=magenta]{\vdots}  (-6.9,11.6);
\draw (-7.4,11.1) node (c0) {}; 
\draw (-6.2,11) node (c00) {}; 
\draw [color=magenta] (c0.north west) rectangle (c00.south east);
\draw (-6.8,11.05) node {\tiny{$\left[\bm{u}_{f_{m-1}}\right]$}};

\path [draw, ->] (-6.1,12) [style={font=\scriptsize}] node[above right=0.5pt]{Sum}-- (-4.9,12) {} ;
\path [draw, ->,dotted] (-6.1,11.6) -- (-4.9,12) {} ;
\path [draw, ->] (-6.1,11.1) -- (-4.9,12) {} ;

\draw (-4.8,12) node (c0) {}; 
\draw (-3.8,11.9) node (c00) {}; 
\draw [color=magenta] (c0.north west) rectangle (c00.south east);
\draw (-4.3,11.95) node {\tiny{$[\texttt{score}_{P[0]}]$}};

\path (-4.2,10.5)  -- node[color=magenta]{$\vdots$}  (-4.2,10.2);
\draw (-2.5,12) node (f0) {}; 
\draw (-0.8,11.9) node (f00) {}; 
\draw [color=magenta] (f0.north west) rectangle (f00.south east);

\path (-1.65,11.5)  -- node[color=magenta]{\vdots}  (-1.65,11.5);
\draw (-2.5,11) node (g0) {}; 
\draw (-0.8,10.9) node (g00) {}; 
\draw [color=magenta] (g0.north west) rectangle (g00.south east);
\node at (-2.3,10.6) [color=magenta,text width = 0.1pt] {{\tiny{\#patients/$2^{12}$}}};

\path [draw, ->] (-3.7,12)  -- (-2.6,12) {} ;
\node at (-3.2,12.2) [style={font=\scriptsize}]{Repack};
\path [draw, ->,dotted] (-3.7,10.4) -- (-2.6,12) {} ;
\path [draw, ->] (-3.7,8.4) -- (-2.6,12) {} ;

\path [draw, ->] (-9,8.6) -- (-8.3,8.6) {} ;
\path [draw, ->,dotted] (-9,8.6) -- (-8.3,8) {} ;
\path [draw, ->] (-9,8.6) -- (-8.3,7.4) {} ;

\draw (-8.1,8.6) node (c0) {}; 
\draw (-8,8.5) node (c00) {}; 
\draw [color=green!50!black!50] (c0.north west) rectangle (c00.south east);
\path (-8.05,8.1)  -- node[color=green!50!black!50]{\vdots}  (-8.05,8.1);
\draw (-8.1,7.6) node (c0) {}; 
\draw (-8,7.5) node (c00) {}; 
\draw [color=green!50!black!50] (c0.north west) rectangle (c00.south east);

\draw (-7.4,8.6) node (c0) {}; 
\draw (-6.2,8.5) node (c00) {}; 
\draw [color=magenta] (c0.north west) rectangle (c00.south east);
\path (-6.7,8.1)  -- node[color=magenta]{\vdots}  (-6.9,8.1);
\draw (-7.4,7.6) node (c0) {}; 
\draw (-6.2,7.5) node (c00) {}; 
\draw [color=magenta] (c0.north west) rectangle (c00.south east);

\draw (-7.7,8.55) node {$\times$};
\draw (-7.7,7.55) node {$\times$};

\path [draw, ->] (-6.1,8.6) -- (-4.9,8.6) {} ;
\path [draw, ->,dotted] (-6.1,8.1) -- (-4.9,8.6) {} ;
\path [draw, ->] (-6.1,7.5) -- (-4.9,8.6) {} ;

\node at (-3,8.2) [color=magenta,text width = 100pt] {{\tiny{one score}}};
\node at (-3.1,8) [color=magenta,text width = 100pt] {{\tiny{per patient}}};

\node at (-5.8,7.2) [color=magenta,text width = 100pt] {{\tiny{encryted $\mathsf{testv}$}}};

\node at (-6.5,7.2) [color=green!50!black!50,text width = 100pt] {{\tiny{one}}};
\node at (-6.5,7) [color=green!50!black!50,text width = 100pt] {{\tiny{coeff}}};

\node at (-12.3,7.7) [color=green!50!black!50,text width = 0.1pt] {{\tiny{\#attributes}}};
\node at (-9.8,7.7) [color=green!50!black!50,text width = 0.1pt] {\tiny{Modified}};
\node at (-9.8,7.4) [color=green!50!black!50,text width = 0.1pt] {\tiny{attributes}};

\draw (-4.8,8.6) node (c0) {}; 
\draw (-3.8,8.5) node (c00) {}; 
\draw [color=magenta] (c0.north west) rectangle (c00.south east);
\draw (-4.3,8.55) node {\tiny{$[\texttt{score}_{\tiny{P[p-1}]}]$}};

\end{tikzpicture}
}
\vspace{1em}
\caption{Illustration of the homomorphic private function evaluation (over selected attributes) phase in TETRIS. At the beginning, the scientist holds the attributes-selection matrix $M$ (in blue) and the scoring functions definitions $f_j$, $j\in[0,m-1]$, and the database owner holds the patients database $P$ (colored in green) represented as a matrix of size $p \times h$.
The scientist first sends the attributes-selection matrix $M$ defined in the plaintext domain and the $m$ attribute scoring functions $f_j$ sent encrypted as a polynomial vector $\RLWE(\bm{u}_{f_{j}})$ (encrypted $\mathsf{testv}$ colored in pink).
For the sake of readability, each encrypted polynomial $\RLWE(\bm{u}_{f_{j}})$ is denoted $[\bm{u}_{f_{j}}]$ in the figure. The results of each scoring function are aggregated together to form an encrypted score for each patient, denoted as $[\texttt{score}_{\tiny{P[j}]}]$, for $j\in [0,p-1]$ (colored in pink).
These intermediate scoring functions are then packed into an encrypted score over a batch of $\# \text{patients}/2^{12}$ (colored in pink).}
\label{fig:usecase-preprocessing}
\end{figure*}

\begin{figure*}
\centering
\scalebox{1.2}{
\begin{tikzpicture}[decoration={brace}][scale=0.5]
\tikzstyle{operation}=[->,>=latex]

\node at (-7,13.5) {\textsc{Homomorphic Operations on the Server's side - Homomorphic thresholds evaluation phase}};

\draw (-12.5,12.5) node (f0) {}; 
\draw (-12.4,11) node (f00) {}; 
\draw [color=magenta] (f0.north west) rectangle (f00.south east);


\node at (-12.9,10.3) [color=magenta][xscale=1.4,yscale=15] {$\{$};

\node at (-13.3,10.5) [color=magenta,text width = 0.1pt] {\rotatebox{90}{\tiny{\#patients/$2^{12}$}}};

\path (-12.45,10.6)  -- node[color=magenta]{\vdots}  (-12.45,10.3);

\draw (-12.5,9.5) node (g0) {}; 
\draw (-12.4,8) node (g00) {}; 
\draw [color=magenta] (g0.north west) rectangle (g00.south east);

\draw (-10.5,11) node (g0) {}; 
\draw (-10.4,9.5) node (g00) {}; 
\draw [color=magenta] (g0.north west) rectangle (g00.south east);
\node at (-11.2,9.2) [color=magenta,text width = 0.1pt] {{\tiny{\#patients/$2^{16}$}}};

\path [draw, ->] (-12.3,11.25) -- (-10.6,10.35) {} ;
\path [draw, ->] (-12.3,9.35) -- (-10.6,10.35) {} ;

\node at (-11.5,11.8) [style={font=\scriptsize}]{Merge};
\node at (-11.4,11.6) [style={font=\scriptsize}]{\&};
\node at (-11.4,11.3) [style={font=\scriptsize}]{Ring Switch};

\draw (-8.5,11.75) node (g0) {}; 
\draw (-8.4,10.25) node (g00) {}; 
\draw [color=magenta] (g0.north west) rectangle (g00.south east);

\draw (-8.5,9.75) node (g0) {}; 
\draw (-8.4,8.25) node (g00) {}; 
\draw [color=magenta] (g0.north west) rectangle (g00.south east);

\node at (-9.2,7.9) [color=magenta,text width = 0.1pt] {{\tiny{\#patients/$2^{16}$}}};

\path [draw, ->] (-10.3,10.25)-- (-8.6,11) {} ;
\path [draw, ->] (-10.3,10.25)-- (-8.6,9) {} ;
\node at (-9.7,10.9) [style={font=\scriptsize}]{\textsf{Half-BTS}};



\draw (-6.5,11.75) node (g0) {}; 
\draw (-6.4,10.25) node (g00) {}; 
\draw [color=magenta] (g0.north west) rectangle (g00.south east);

\draw (-6.5,9.75) node (g0) {}; 
\draw (-6.4,8.25) node (g00) {}; 
\draw [color=magenta] (g0.north west) rectangle (g00.south east);

\node at (-7.2,7.9) [color=magenta,text width = 0.1pt] {{\tiny{\#patients/$2^{16}$}}};

\path [draw, ->] (-8.3,11) -- (-6.6,11) {} ;
\node at (-7.4,11.2) [style={font=\scriptsize}]{Thres1-p};

\path [draw, ->] (-8.3,9) -- (-6.6,9) {} ;
\node at (-7.4,9.2) [style={font=\scriptsize}]{Thres1-p};

\draw (-4.5,11) node (g0) {}; 
\draw (-4.4,9.5) node (g00) {}; 
\draw [color=magenta] (g0.north west) rectangle (g00.south east);
\node at (-5.2,9.2) [color=magenta,text width = 0.1pt] {{\tiny{\#patients/$2^{16}$}}};

\path [draw, ->] (-6.3,11) -- (-4.6,10.25) {} ;
\path [draw, ->] (-6.3,9) -- (-4.6,10.25) {} ;
\node at (-5.3,10.9) [style={font=\scriptsize}]{Sum};

\draw (-6.5,9.75) node (g0) {}; 
\draw (-6.4,8.25) node (g00) {}; 
\draw [color=magenta] (g0.north west) rectangle (g00.south east);

\draw (-2.5,11) node (g0) {}; 
\draw (-2.4,9.5) node (g00) {}; 
\draw [color=magenta] (g0.north west) rectangle (g00.south east);
\node at (-3,9.2) [color=magenta,text width = 0.1pt] {{\tiny{\#patients/$2^{16}$}}};


\draw (-0.8,10.3) node (g0) {}; 
\draw (-0.7,10.2) node (g00) {}; 
\draw [color=magenta] (g0.north west) rectangle (g00.south east);
\node at (-0.8,9.9) [color=magenta,text width = 0.1pt] {{\tiny{$1$}}};

\node at (-3.4,10.5) [style={font=\scriptsize}]{Inner Sum};

\node at (-1.6,10.5) [style={font=\scriptsize}]{Thres2-p};

\path [draw, ->] (-2.3,10.25) -- (-0.9,10.25) {} ;

\path [draw, ->] (-4.3,10.25) -- (-2.6,10.25) {} ;

\end{tikzpicture}
}
\vspace{-5em}
\caption{Illustration of the homomorphic thresholds evaluation phase in TETRIS. This phase is processed on the database owner side. At the beginning, he holds a set of $\#\text{patients}/2^{12}$ ciphertexts which are merged into a ciphertext in a small ring of dimension $N=2^{12}$ and switched to a larger ring of $2^{16}$ before entering into the CKKS bootstrapping. The output is two ciphertexts, each encrypting a vector of $N/2=2^{15}$ encoded scores for the $\#\text{patients}/2^{16}$ patients. Then the database owner evaluates a first local threshold, denoted as $\mathsf{thres1}\text{-}\mathsf{p}$ in the figure and aggregates all the encrypted results via homomorphic summation. The database owner then applies the second private global threshold, denoted as $\mathsf{thres1}\text{-}\mathsf{p}$ in the figure and outputs the encrypted result.}
\label{fig:usecase-computations}
\end{figure*}
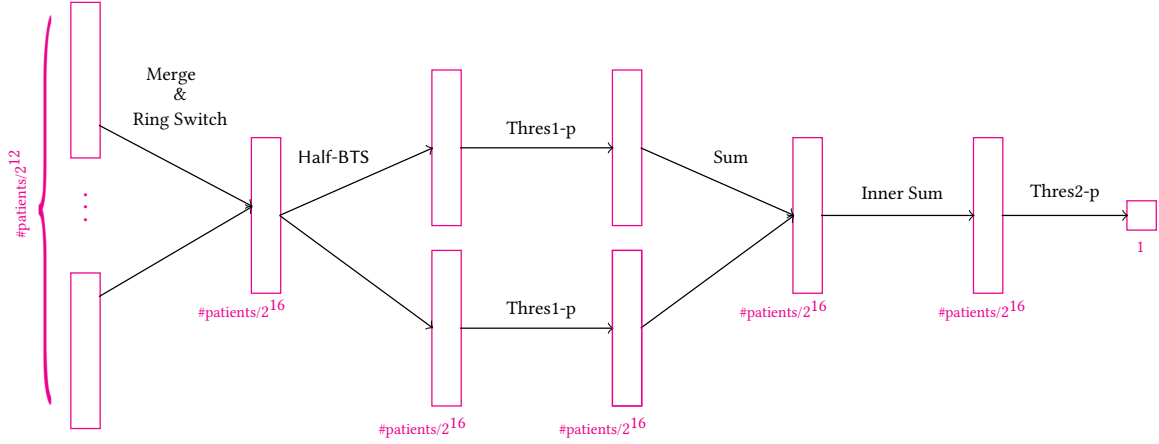

We detail in Figure~\ref{fig:usecase-preprocessing} and Figure~\ref{fig:usecase-computations} the different steps of the homomorphic private function evaluation phase and the homomorphic thresholds evaluation phase processed by the database owner. 

\vspace{-1.5em}

\bigskip 

\subsubsection{Circuit Structure}
The full pipeline of the evaluations circuits is summarized in Algorithm~\ref{alg:privatedatabaseexploration}. 
Before the protocol begins, the scientist holds the attributes-selection matrix $M$ and the scoring functions and the database owner holds the patients’ database.

We review the evaluation process below:
\begin{enumerate}[label=(\roman*)]
\item The scientist generates:
\begin{itemize}
\item 
the FHE secret and the necessary evaluation keys, including the keyswitching key and the bootstrapping key;
\item
the $m$ encrypted scoring functions $\left[\RLWE(\bm{u}_{f_{0}}), \dots,\right.$\\ $\left. \RLWE(\bm{u}_{f_{m-1}})\right]$;
\item
the local encrypted threshold parameter $\RLWE(t_0)$ and normalization factor $\RLWE\left(\frac{1}{\sum_{j=0}^{m-1} \max(f_{j})}\right)$;
\item
the global encrypted threshold parameter $\RLWE(t_1)$ and normalization factor $\RLWE\left(\frac{1}{\sum_{j=0}^{m-1} \max(f_{j})}\right)$.
\end{itemize}
\item
The scientist sends evaluation keys together with the encrypted values to the database owner;
\item 
the database owner applies the attributes-selection matrix to $P$ to obtain a new patients’ database $P'$ and evaluates:\\ $\RLWE(\textsf{score}_{P'[i]})= \sum_{j=0}^{m-1} \RLWE(\bm{u}_{f_{j}}) \cdot X^{P'[i][j]}$ for $i\in [0,p-1]$.
It sets $\textsf{ct}_i:=\RLWE(\textsf{score}_{P'[i]})$; 
\item
the database owner applies the homomorphic ring packing to the $p$ low degree $\RLWE$ ciphertexts;
\item
the database owner applies the homomorphic ring merging operation to switch to a large degree $\RLWE$ ciphertexts;
\item
the database owner applies the scheme switching operation via the partial bootstrapping $\texttt{Half-BTS}$ operation to obtain CKKS ciphertexts;
\item
the database owner evaluates the local threshold homomorphically and aggregates all the values;
\item
the database owner evaluates the final threshold homomorphically and sends back the result to the scientist.
\end{enumerate}

\begin{algorithm}[h]
\caption{Private Database Exploration}
\label{alg:privatedatabaseexploration}
\begin{algorithmic}[1]
\State \underline{Inputs:} 
\STATEx A database $P$ given as a $p\times h$ matrix
\STATEx- {$M$ an $h\times m$ attributes-selection matrix}
\STATEx- {$[\RLWE(\bm{u}_{f_{0}}), \dots, \RLWE(\bm{u}_{f_{m-1}})] \in \left((\ring{}{n}{Q_{0}})^{2}\right)^m$ a vector of encrypted scoring function}
\STATEx- $\RLWE\left(\tfrac{1}{\sum\max(f_{i})}\right) \in (\ring{}{N}{Q_{\ell}})^{2}$
\STATEx- $\RLWE(t_{0}), \RLWE(t_{1}) \in (\ring{}{N}{Q_{\ell}})^{2}$, two private threshold values sent encrypted by the scientist to the database owner.
\STATEx
\State \underline{Output:} $\RLWE(b)$ where $b=1$ if at least $t_{1}$ entries of $P$ produce individual an score of at least $t_{0}$, else $b=0$.
\STATEx
\State $P' \leftarrow P \times M$ \tcp{Attributes selection}
\State $\textsf{ctsN12} \leftarrow [\emptyset]$
\State $\textbf{for}\ i \gets 0$ \textbf{to} $p-1$ \textbf{do}
\State \qquad $\textsf{ct}_i\leftarrow \sum_{j=0}^{m-1} \RLWE(\bm{u}_{f_{j}}) \cdot X^{P'[i][j]} \in (\ring{}{n}{Q_{0}})^{2}$
\State $\textsf{ctsN12} \leftarrow \texttt{Repack}(\textsf{ct}_0,\cdots,\textsf{ct}_{p-1})$ 
\STATEx
\State $\textsf{ctsN16} \leftarrow \texttt{Merge}(\textsf{ctsN12})$ 
\hspace{6cm} \tcp{ Ring merging \& ring switching \hspace{-1cm} \\ $\ring{}{n}{Q_{0}}\rightarrow\ring{}{N}{Q_{0}}$}
\STATEx
\State $\textsf{ctScore}\leftarrow \emptyset$
\State \textsf{for} $i\leftarrow 0$ \textbf{to} $\lceil (p(n/N)\rceil$ \textbf{do}
\State \qquad $\left(\textsf{ctL}, \textsf{ctR} \right)\leftarrow \texttt{Half-BTS}(\textsf{ctsN16}[i])$ \tcp{\small{$\ring{}{N}{Q_{0}}\rightarrow \ring{}{N}{Q_{L}}$}}
\State \qquad $\textsf{ctScore} \leftarrow \textsf{ctScore} + \texttt{step}\left((\textsf{ctL} - \RLWE(t_{0}))\cdot\RLWE\left(\tfrac{1}{\sum\max(f_{i})}\right)\right)$
\State \qquad $\textsf{ctScore} \leftarrow \textsf{ctScore} + \texttt{step}\left((\textsf{ctR} - \RLWE(t_{0}))\cdot\RLWE\left(\tfrac{1}{\sum\max(f_{i})}\right)\right)$\\
\State return $\texttt{step}\left((\texttt{InnerSum}(\textsf{ctScore}) - \RLWE(t_{1})) \cdot p^{-1} \right)$
\end{algorithmic}
\end{algorithm}

\subsection{Extending Algorithm~\ref{alg:privatedatabaseexploration} to Partitioned Databases}

In the case of a horizontally or vertically partitioned database, Algorithm~\ref{alg:privatedatabaseexploration} either needs trivial or no modification since we are dealing with counts and only one of the database owner is required to download the bootstrapping keys (7.4GB), while all others only need to download the attributes-selection plaintext matrix, the HE-based PFE and the repacking keys (a few MB).
In both cases, all database owners can proceed without modification with the PFE evaluation and repacking until line 7 of Algorithm~\ref{alg:privatedatabaseexploration} and then send their repacked ciphertexts to the database owner holding the bootstrapping keys.
Note that repacked ciphertexts after the PFE evaluation (line 7 of Algorithm~\ref{alg:privatedatabaseexploration}) have a small expansion ratio and can store up $2^{12}$ scores each, for a ciphertext size of 65KB, meaning that transmitting the $2^{19}$ scores at this stage would only require $2^{19}/2^{12}\cdot65536 \approx 8$MB to transmit $p$ scores under this form.
\subsubsection{Horizontally Partitioned Database}
Once the database owner holding the bootstrapping keys has received all ciphertexts from the others, it can proceed from line $8$ of Algorithm~\ref{alg:privatedatabaseexploration} until the end and without additional modification.
If the respective value $p^{-1}$ of each database owner cannot be shared, it can be homomorphically computed by multiplying the encryptions of their respective $p^{-1}$.
\subsubsection{Vertically Partitioned Database}
Once the database owner holding the bootstrapping keys has received all ciphertexts from other database owners, it must aggregate them before being able to proceed starting from line $8$ of Algorithm~\ref{alg:privatedatabaseexploration} until the end and without additional modification.
In this scenario the value $p$ is the same for all database owners and thus known by all.

\section{Implementation \& Performance}
\label{section:implem}
\subsection{Cryptographic Parameters}
The private database exploration circuit uses three sets of cryptographic parameters:
\begin{itemize}[label={$\bullet$}]
\item \textbf{Set I:} The parameters used for the PFE and for the Ring-Packing and Merging.
\item \textbf{Set II:} The parameters used for the private thresholds evaluation.
\item \textbf{Set III:} The parameters used for the bootstrapping circuit, which contain as a subset the parameters of \textbf{Set II}.
\end{itemize}

The details of these parameters can be found in Table~\ref{tab:praticalcalse:parameters}.

\begin{table*}
\centering
\begin{tabular}{|c|c|c|c|c|c|c|c|}
\hline
Set & $\lambda$ &$\log(N)$ & $\log(QP)$ & $\log(Q)$ & $\log(P)$ & \texttt{base2} &$h$\\
\hline
I & 128 & 12 & 109 & 55 & 54 & 30 & 2N/3\\
\hline
II & $256$ & 16 & 577 & $55 + 8\cdot45$ & $3\cdot 54$ & - & 192\\
\hline
III & 128 & 16 & 1541 & $55+8\cdot45+3\cdot39+8\cdot60+4\cdot56$ & $5\cdot61$ & - & 192\\
\hline
\end{tabular}
\caption{Cryptographic Parameters Sets. \textsf{base2} refers to an additional base 2 decomposition during the key-switching, which is required since $P$ cannot be set to be larger without reducing the security parameters.}
\label{tab:praticalcalse:parameters}
\end{table*}

\begin{table}
\centering
\begin{tabular}{|c|c|c|c|}
\hline
& $\alpha$ & $\beta$ & degrees \\
\hline
Threshold1-p & 8 & 12 & [15, 15, 15] \\
\hline
Threshold2-p & 16 & 20 & [15, 15, 15, 15, 15] \\
\hline
\end{tabular}
\vspace{0.3cm}
\caption{Parameters of the threshold functions: $\alpha$ parametrizes the sensitivity of the threshold, i.e. the interval $[-1, -2^{-\alpha}]\cup[2^{-\alpha}, 1]$ in which the function returns a correct value and $\beta=-\log_{2}(|b-\round{b}|)$, the precision of the output bit $b$}.
\label{tab:threshold}

\end{table}
\subsection{Other Parameters}
Besides the cryptographic parameters, we need to define parameters for the bootstrapping and private threshold functions.

The parametrization of the bootstrapping circuit, uses the default parameters~\cite{bossuat2022} of the Lattigo library~\cite{lattigo}: a ring degree of $N=2^{16}$, a circuit-depth of $k=15$ (for a total modulus consumption of 821 bits), a ternary secret with Hamming-weight $h=192$, an ephemeral secret with Hamming-weight $\tilde{h}=32$, which provide a failure probability of $2^{-138.7}$ and a precision of $27.25$ bits for a message $\mathbb{C}^{32768}$ with the real and imaginary part uniformly distributed in $[-1, 1]$. The parametrization of the private threshold functions can be found in Table~\ref{tab:threshold}.

\subsection{Scientist's Data} The scientist's data sent to the database owner comprises the evaluation keys as well as the encrypted test vectors and encrypted threshold values.
The size for each of these objects can be found in 
Table~\ref{tab:praticalcase_clientdata}.

\begin{table}
\centering
\begin{tabular}{|c|c|}
\hline
Set & Size [MB] \\
\hline
Ring Packing Keys & 3.25\\
\hline
Ring Merging Keys & 15.0\\
\hline
Bootstrapping Keys & 7395\\
\hline
$\RLWE(f_i)$ & 1.04\\
\hline
$\RLWE(t_i)$ & 4.5 + 2.25\\
\hline
\hline
Total & $\approx 7421$\\
\hline
\end{tabular}
\caption{Scientist's Data}
\label{tab:praticalcase_clientdata}
\end{table}


\subsection{Performance} 
We implemented Algorithm~\ref{alg:privatedatabaseexploration} using the Lattigo library~\cite{lattigo}. The code can be found in the following repository~\url{https://github.com/izama/private-database-exploration}. All benchmarks were conducted on the following hardware: Go 1.21, Windows 11, i9-12900K, 32GB DDR4, single threaded. 
The code is available in the supplementary materials. Table~\ref{tab:practicalcase_timings} reports the timings for a database of $2^{19}$ entries with $16$ attributes after the evaluation of the attributes-selection matrix.
Three out of the 6 values reported are independent of the number of entries in the database:
(i) the scientist generation which consist in the evaluation-keys generation, (ii) the database owner generation which consists in the instantiation of the bootstrapping evaluator (generation of the plaintext matrices for the \texttt{CoeffsToSlots} and \texttt{SlotsToCoeffs} steps) and (iii) the second private threshold which is evaluated on the final count.
The three other timings amortize to $1.85$ms per entry or $32555$ entries per minute, single threaded.

\begin{table}[h]
\centering
\begin{tabular}{|l|c|c|}
\hline
Operation & Total & Amortized \\
\hline
Scientist Generation & $24.4$s & $46\mu$s \\
\hline
Database Owner Generation & $10.9$s & $20\mu$s \\
\hline
Private Function \& Packing & $466.5$s & $888\mu$s\\
\hline
Half-BTS & $73.4$s & $140\mu$s\\
\hline
Private Threshold 1 & $384.1$s & $732\mu$s\\
\hline
Private Threshold 2 & $58.7$s & $111\mu$\\
\hline
\hline
Total & 1018.0 & $1.941$ms\\
\hline
\end{tabular}
\caption{Timings for a database of $2^{19}$ entries and $16$ attributes, single threaded.}
\label{tab:practicalcase_timings}
\end{table}

\subsection{Comparison with Sending the Database Encrypted.}
\label{sec:comparison}
Using an HE-based PFE greatly outperforms the alternative, which is to send the database encrypted to the scientist.
Although this would remove the need for the scientist to transmit the 7.4GB bootstrapping keys to the server, encrypting a database of $p$ patients and $h$ features has an expansion ratio that is $\mathcal{O}(|\textsf{Dom}_{f}|\cdot|\log(Q)|)$ (our HE-based PFE requires one-hot encoding of the inputs).
In our use case we have $|\textsf{Dom}_{f}| = 2^{12}$, $p=2^{19}$, $h=16$ and $\log(Q)=55$ (8 bytes). 
As such transmitting the same database in an encrypted form to the scientist would require a communication in the order of $8\cdot(2^{12}\cdot2^{19}\cdot16) = 64\text{GB}$.
Finally, it is noteworthy to point out that this 7.4GB of keys is independent of the circuit and database size and that sending the database in encrypted form to the scientist would leak its size, which remains hidden when using the PFE approach.

\subsection{Comparison with PFE}\label{sec:uc_vs_fhe}
Recalling Section~\ref{section:intro}, PFE originates from UC-based constructions, first introduced by Valiant~\cite{Valiant76}.
A recent work of Holz et al.~\cite{ESORICS:HKRS20} compares the efficiency of different UC-based approaches PFE with Homomorphic-based approaches PFE when using Elliptic curve ElGamal~\cite{ElGamal85}, the Brakerski/Fan-Vercauteren (BFV) scheme~\cite{Fan2012Bfv} and Damg{\aa}rd-Jurik-Nielsen~\cite{DamJurNie10}. Similar to Holz et al., we split our protocol in different phases with a first phase where the party owning the private data can learn either the size of the function or some part of the function such that the global homomorphic evaluation  becomes faster.
In our framework during the homomorphic private function evaluation phase, the database owner evaluates a private function. In our case, the private function  is a scoring function but the method we propose can be used to evaluate an arbitrary function. 
Making a direct and fair comparison with their work is difficult because of the hardware difference (32 Core @ 2.8Ghz vs. ours which is single thread @ 4.9Ghz) and because they express the complexity of their circuit in term of the number of boolean gates, but we can make a rough estimate.
Our HE-based PFE can be expressed as the evaluation of a polynomial of $d=2^{12}$ coefficients over GF($2^{12}$), which requires as many multiplications and additions over GF($2^{12}$).
One addition over GF($2^{12}$) requires $12$ XOR gates while a multiplication by $X$ requires $4$ XOR gates (modular reduction by a primitive polynomial of Hamming weight $5$).
As such the multiplication of an unknown field element by a fixed field element $a$ requires $\approx\textsf{HW}(a)\cdot(12 + 4)$ XOR gates.
Therefore, the circuit size would be in the order of $10^{5}$ to $10^{6}$ gates.
For $10^{6}$ gates, Holz et al.~\cite[Table 1]{ESORICS:HKRS20} report a setup of between 300MB with EC ElGamal (smallest) to 2.3GB UC per function (largest), which would amount to 4.8GB for EC ElGamal and 37GB for UC, for 16 different functions.
Discarding the setup, since with our approach it is negligible (in the order of a few ms to generate RLWE ciphertexts) and becomes negligible for all approaches when the PFE is evaluated over many points,~\cite[Table 2]{ESORICS:HKRS20} reports an online time of $1.15$ sec per evaluation with BFV (best) to 19 sec with UC (worst) per function evaluation. 
In our work we only need 0.125MB per function as well as 3.25MB for the repacking keys and have an amortize time of $888/16 = 55\mu\text{s}$ per function evaluation (which is $\approx20'000\times$ faster).

\section{Conclusion}
\label{section:conclusion}
In this paper, we present TETRIS, a practical solution that allows a scientist to explore a large dataset of patients’ records, while maintaining both the privacy of the patients and exploration criteria.
Our solution is also communication efficient in the sense that the database does not need to be sent at any point of the computation.
We adapt efficient HE-based techniques for arbitrary functions over large domains with the CKKS bootstrapping to support customized and efficient homomorphic circuit compositions.
Our experimental results show that a database with $2^{19}$ entries and $16$ features has an amortized total homomorphic processing time of around $1.9$ms per entry.
Our protocol can be easily extended to support multiple databases, either horizontally or vertically partitioned. 

\section*{Acknowledgments}
This research received no specific grant from any funding agency in the public, commercial, or not-for-profit sectors.

\bibliographystyle{alpha}

\bibliography{bib/cryptobib/abbrev3,bib/cryptobib/main}

\section*{Appendix}

\section{Additional preliminaries}
\label{appendix:preliminaries}

\subsection{Plaintext Domains}\label{appendix:plaintextdomains}
Throughout this work we will distinguish between two plaintext domains: in the ring (a.k.a in the \emph{coefficients}) and in the canonical embedding (a.k.a in the \emph{slots}).

\subsubsection{In the ring}
Messages are said to be in the \emph{coefficients} if their encoding provides an arithmetic that follows the regular arithmetic of $\mathcal{R}^{N}$ (see Section~\ref{section:cyclotomicrings}).
Encoding a message $m\in\mathbb{R}^{n}$ for $n \leq N$ on $\mathcal{R}^{N}$ (and its inverse operation) is done with the following steps:
\begin{equation*}
\mathbb{R}^{n} \xleftrightarrow[(1)]{} \mathbb{R}[Y] \xleftrightarrow[1/\Delta]{\lfloor\Delta\cdot\rceil} \mathbb{Z}[Y] \xleftrightarrow[(2)]{} \ring{[X]}{N}{},
\end{equation*}

for $Y = X^{N/n}$.

\begin{itemize}
\item $(1)$: the mapping $\mathbb{R}^{n} \rightarrow \mathbb{R}[Y]$ is defined as interpreting a vector in $\mathbb{R}^{n}$ as a polynomial in $Y$ of $n$ real coefficients.
\item $(2)$: the mapping $\mathbb{Z}[Y] \rightarrow \ring{[X]}{N}{}$ is defined as taking an integer polynomial in $Y$ and embedding it in $\ring{[X]}{N}{}$ by applying the change of variable $Y \rightarrow X^{N/n}$.
\end{itemize}

The scaling factor $\Delta$ is used to map $\mathbb{R}$ to $\mathbb{Z}$ via fixed-point arithmetic.
Naturally, it will grow from $\Delta$ to $\Delta^{2}$ after a multiplication and its magnitude is managed by applying $\lfloor1/\Delta\rceil$ on the coefficients and the modulus of $\ring{[X]}{N}{}$.
In the following, we will denote $\lll_{k}$ for the cyclic rotation by $k$ positions to the left of a vector or polynomial. As $\mathbb{Z}^{*}_{2N}$ is generated by $-1$ and $5$ (see Section~\ref{section:cyclotomicrings}), we will denote $\phi_{(k_0, k_1)}$ the automorphism which maps $X^i$ to $X^{i\cdot (-1)^{k_0} (5)^{k_1}}$.

\subsubsection{In the Canonical Embedding}

Messages are said to be in the \emph{slots} if their plaintext arithmetic, when doing operation in $\mathcal{R}$, behaves as single-instruction-multiple-data (SIMD) over $\mathbb{C}^{n}$.

Let $n$ be a power-of-two integer such that $1 \leq n < N$ and $\psi = e^{i\pi/{n}}$ be a $2n$-th primitive root of unity.
Since $(-1)^{k_{0}}5^{k_1}$ for $0 \leq k_0 < 2$ and $0 \leq k_1 < n/2$ spans $\mathbb{Z}_{2n}^{*}$, we have $\{\psi^{5^{k}}, \overline{\psi^{5^{k}}}, 0\leq k < n/2\}$ the set of all $2n$-th primitive roots of unity, which form an orthogonal basis over $\mathbb{C}^{n}$.

The decoding complex discrete Fourier transform of dimension $n$ ($\textsf{DFT}_{n}$) is characterized by the $n\times n$ special Fourier matrix $\text{SF}_{n}[j][k] = \psi^{j5^{k}}$ (up to a bit-reversal permutation) for $0\leq j, k < n$, with its inverse ($\textsf{DFT}^{-1}_{n}$), the encoding matrix, being $\text{SF}^{-1}_{n} = \frac{1}{n}\overline{\text{SF}}^{T}_{n}$ due to the orthogonality of the basis.
For $Y = X^{N/n}$, encoding a vector $\bm{m}\in\mathbb{C}^{n}$ on $\ring{[X]}{N}{}$ (and its inverse operation) is similar to encoding in the \emph{coefficients} but with a pre-processing step involving the evaluation of $\textsf{DFT}^{-1}_{n}$ (or $\textsf{DFT}_{n}$ for the decoding) on $\bm{m}$:
\begin{equation*}
\mathbb{C}^{n} \xleftrightarrow[\textsf{DFT}_{n}]{\textsf{DFT}^{-1}_{n}} \mathbb{C}^{n} \xleftrightarrow[(1)]{} \mathbb{R}^{2n} \xleftrightarrow[(2)]{} \mathbb{R}[Y] \xleftrightarrow[1/\Delta]{\lfloor\Delta\cdot\rceil} \mathbb{Z}[Y] \xleftrightarrow[(3)]{} \ring{[X]}{N}{}.
\end{equation*}

\begin{itemize}
\item $(1)$: the mapping $\mathbb{C}^{n} \rightarrow \mathbb{R}^{2n}$ maps a complex vector of size $n$ to a real vector of size $2n$ and is defined as $\Re(\bm{m})||\Im(\bm{m})$ (its inverse being defined as $\bm{m}[:n] + i\cdot \bm{m}[n:]$).
\item $(2)$: the mapping $\mathbb{R}^{2n} \rightarrow \mathbb{R}[Y]$ is defined as interpreting the vector $\bm{m}$ as a polynomial in $Y$ of $2n$ real coefficients.
\item $(3)$: the mapping $\mathbb{R}[Y] \rightarrow \ring{[X]}{N}{}$ is defined as taking a polynomial in $Y$ and embedding it in $\ring{[X]}{N}{}$ by applying the change of variable $Y \rightarrow X^{N/n}$.
\end{itemize}

We will identify messages in the \emph{slots} with the brackets $\langle \cdot \rangle$ (else they are assumed to be in the \emph{coefficients}) and for any two vectors $\bm{m}_{0},\bm{m}_{1} \in \mathbb{C}^{n}$, the following holds due to the convolution property of the complex \textsf{DFT}:
\begin{itemize}
\item \textsf{Addition}: $\langle \bm{m}_{0}\rangle + \langle \bm{m}_{1} \rangle = \langle \bm{m}_{0} + \bm{m}_{1} \rangle$
\item \textsf{Multiplication}: $\langle \bm{m}_{0} \rangle * \langle \bm{m}_{1} \rangle = \langle \bm{m}_{0} \cdot \bm{m}_{1} \rangle$
\end{itemize}

\noindent Additionally, automorphisms on messages in the \emph{slots} act as vector operations in the following way:
\begin{itemize}
\item \textsf{Rotation}: $\phi_{(0, k)}(\langle \bm{m}\rangle) = \langle \bm{m}\lll_{k} \rangle$
\item \textsf{Conjugation}: $\phi_{(1, 0)}(\langle \bm{m}\rangle) = \langle \overline{\bm{m}}\rangle$ 
\end{itemize}

The same scaling factor management as the \emph{coefficient} encoding is required after multiplication.
For the rest of this work, we will denote $\round{\Delta\tau_{n}(m)}$ the process of encoding $\bm{m}\in\mathbb{C}^{n}$ on $\ring{[X]}{N}{}$ and $\tau^{-1}(\Delta^{-1}m)$ its inverse.

\subsection{Advanced Homomorphic Operations}\label{appendix:advanced_operations}
We detailed below how homomorphic operations are performed using key-swicthing:
\begin{itemize}
\item \textsf{Relinearization}: the multiplication between two $\RLWE$ ciphertexts $\prlwe{}{s}{}$ returns a new $\RLWE$ ciphertext $\prlwe{2}{s}{}$, and the degree will further increase if we perform additional multiplication with non-plaintext operands.
To ensure the compactness of the multiplication, the key-switching operation is used to homomorphically re-enrypt $\prlwe{2}{s}{}$ to $\prlwe{1}{s}{}$ by evaluating $\textsf{SwitchKey}(\prlwe{[2]}{s^{2}}{}, \textsf{swk}_{s^{2}\rightarrow s}) + \prlwe{[0:1]}{s}{}$;\\
\item \textsf{Automorphism}: an automorphism $\phi$ on an $\RLWE$ ciphertext $\prlwe{}{s}{}$ maps it to a new RLWE ciphertext $\prlwe{}{\phi(s)}{}$.
The key-switching operation enables to re-encrypt $\prlwe{}{\phi(s)}{}$ to $\prlwe{}{s}{}$ by evaluating $\textsf{SwitchKey}(\prlwe{[1]}{\phi(s)}{}, \textsf{swk}_{\phi(s)\rightarrow s}) + \prlwe{[0]}{\phi(s)}{}$;\\
\item \textsf{Ring Splitting}: it is possible to split an RLWE ciphertext $\prlwe{}{s}{} + m$ with $m,s\in\ring{}{N}{}$ into two ciphertexts of half the dimension $(\prlwe{}{s'}{} + m_{0}, \prlwe{}{s'}{} + m_{1})$ with $m_{0}, m_{1}, s'\in\ring{}{N/2}{}$ such that $m(X) = m_{0}(Y) + X*m_{1}(Y)$ for $Y = X^{2}$, by evaluating the following: $d = \prlwe{}{s'(Y)}{} + m = \textsf{SwitchKey}\left(\prlwe{[1]}{s}{},\right.$\\ $\left.\textsf{swk}_{s\rightarrow s'(Y)}\right) + \prlwe{[0]}{s}{} + m$ and returning $(d(Y \rightarrow X), (X*d)(Y \rightarrow X))$, where $Y\rightarrow X$ denotes the change of variable from $Y$ to $X$. This can be generalized to $Y = X^{2^{i}}$;\\
\item \textsf{Ring Merging}: it is possible to merge two RLWE ciphertexts $d_{0} = \prlwe{}{s'}{} + m_{0}$ and $d_{1} = \prlwe{}{s'}{} + m_{1}$ with $m_{0}, m_{1}, s'\in\ring{}{N/2}{}$ into an RLWE ciphertext of twice the dimension $\prlwe{}{s}{} + m$ with $m,s\in\ring{}{N}{}$ such that $m(X) = m_{0}(Y) + X*m_{1}(Y)$ for $Y = X^{2}$ by evaluating the following: $\prlwe{}{s'(Y)}{} + m = d_0(Y) + X*d_1(Y)$ and returning $\prlwe{}{s}{} + m = \textsf{SwitchKey}\left(\prlwe{[1]}{s'(Y)}{}\right.,$\\$\left. \textsf{swk}_{s'(Y)\rightarrow s}\right) +\prlwe{[0]}{s'(Y)}{} + m$. This can be generalized to $Y = X^{2^{i}}$.\\
\end{itemize}

\subsection{Bootstrapping for Approximate Homomorphic Encryption}\label{appendix:bootstrapping}
In this section, we recall the CKKS bootstrapping fours steps:

\noindent \textsf{ModRaise}: the ciphertext $\prlwe{}{s}{q}+m(Y)$ is expressed as modulo $Q$.
This returns a new ciphertext $\prlwe{}{s}{Q}+ m'(X)$ where $m'(X) = q\cdot I(X) + m(Y)$.
The unwanted integer polynomial $q\cdot I(X)$ comes from the absence of a reduction modulo $q$ during the decryption process with $||I(X)||$ being $\mathcal{O}(\sqrt{h})$.
If $2n \neq N$, then $Y \neq X$ and $I(X)$ is not a polynomial in $Y$.
In other words, we have multiples of $q$ in the coefficients $X$ that are not multiples of $N/2n$.
In this case, we can map $q\cdot I(X) + m(Y)$ to $q\cdot I(Y) + m(Y)$ by evaluating a trace-like map~\cite{EC:CHKKS18} that maps $X\rightarrow Y$ making coefficients in $X$ whose degree is not a multiple of $N/2n$ vanish.
This map can be efficiently evaluated with $\log(N/2n)$ automorphisms~\cite{EC:CHKKS18} and doing so reduces the complexity of the next step.
The remaining steps of the bootstrapping are aimed at removing $q \cdot I(Y)$ by homomorphically evaluating a coefficient-wise modular reduction by $q$ on $m'$.\\

\noindent \textsf{CoeffsToSlots}: $m'(Y)$ can be seen as a fresh message in the \emph{coefficients} and SIMD evaluation, it needs to be encoded in the \emph{slots}.
To do so, this step homomorphically evaluates $\tau_{n}(m'(Y))$ returning $\langle m'(Y)\rangle$;\\

\noindent \textsf{EvalMod}: a polynomial approximation of the modular reduction by $q$ is homomorphically evaluated on $\langle m'(Y)\rangle$, making the polynomial $I(Y)$ vanish;\\

\noindent \textsf{SlotsToCoeffs}: the inverse of the canonical embedding, $\tau^{-1}_{n}$, is evaluated on $\langle m'(Y)\rangle$ and a close approximation of original message $m(Y)$ minus the unwanted polynomial is retrieved.
After this last step the ciphertext has modulus $Q' > \Delta^{2} > q$ and we can evaluate further operations, until the ciphertext becomes \emph{exhausted} again and a new bootstrapping is needed.\\

Note that a ciphertext at modulus $Q' < Q$ is returned because the \emph{bootstrapping} circuit consumes part of the original modulus $Q$ as it requires to call rescaling operations during the \textsf{CoeffsToSlots}, \textsf{EvalMod} and \textsf{SlotsToCoeffs} steps.

\subsection{\textsf{CoeffsToSlots} and \textsf{SlotsToCoeffs}}
\label{appendix:coeffslots}
The goal of the \textsf{CoeffsToSlots} and \textsf{SlotsToCoeffs} steps is to homomorphically evaluate $\tau_{n}$ on a message $m$.
Since the encoding and decoding discrete Fourier transforms are fully defined by the $n\times n$ matrices $\text{SF}_{n}$ and $\text{SF}^{-1}_{n}$ respectively (see Section~\ref{appendix:plaintextdomains} which details the plaintext domain spaces), its homomorphic evaluation can be expressed in terms of plaintext matrix-vector products:
\begin{enumerate}
\item \textsf{CoeffsToSlots}$(m_{rev})\rightarrow \langle m \rangle: m_{rev} \times \text{SF}^{-1}_{n}$;
\item \textsf{SlotsToCoeffs}$(\langle m \rangle)\rightarrow m_{rev}: \langle m \rangle\times\text{SF}_{n} $,
\end{enumerate}

\noindent where the suffix $_{rev}$ denotes the bit-reversal permutation.

In their initial bootstrapping proposal, Cheon et al. \cite{EC:CHKKS18} homomorphically compute the DFT as a single matrix-vector product in $\mathcal{O}(\sqrt{n})$ rotations and depth 1, by using the baby-step giant-step (BSGS) approach of Halevi and Shoup \cite{Halevi2015HelibBoot}.
To further reduce the complexity, two recent works from Cheon et al. \cite{Cheon2018FasterDFT} and Chen et al. \cite{Chen2019ImprovedBootstrapping} exploit the structure of the equivalent FFT algorithm by recursively merging its iterations, reducing the complexity to $\mathcal{O}(\sqrt{r}\log_{r}(n))$ rotations at the cost of increasing the depth to $\mathcal{O}(\log_{r}(n))$, for $r$ a power-of-two radix between $2$ and $n$.

Note that this FFT factorization does not extend to the bit-reversal permutation as it cannot be factorized, thus it is omitted in practice, and we might require to adapt the following homomorphic operations to bit-reversal inputs.

\end{document}